\documentclass[12pt]{article}\usepackage{knitr}
\usepackage{natbib}
\usepackage{amsmath}
\usepackage{amssymb}
\usepackage{bm}
\usepackage{acronym}
\usepackage[super]{nth}
\usepackage[margin=1in]{geometry}
\usepackage{caption} 

\newcommand{\Cpp}{C\nolinebreak\hspace{-.05em}\raisebox{.4ex}{\tiny\bf +}\nolinebreak\hspace{-.10em}\raisebox{.4ex}{\tiny\bf +}}

\newcommand\mat[1]{\mathbf{#1}}
\newcommand\bigO[1]{\mathcal{O}\left(#1\right)}
\renewcommand\vec{\bm}
\newcommand\der{\operatorname{d\!}{}}

\DeclareMathOperator\Prob{P}

\DeclareMathOperator\diag{diag}
\DeclareMathOperator*{\argmax}{arg\,max}

\acrodef      {VA}{variational approximation}
\acrodefplural{VA}{variational approximations}

\acrodef      {GLMM}{generalized linear mixed model}
\acrodefplural{GLMM}{generalized linear mixed models}

\acrodef      {CDF}{cumulative density function}
\acrodefplural{CDF}{cumulative density functions}

\acrodef      {GSM}{generalized survival model}
\acrodefplural{GSM}{generalized survival models}

\acrodef      {PDF}{probability density function}

\acrodef      {MC}{Monte Carlo}

\acrodef      {KL}{Kullback–Leibler}

\acrodef      {GHQ}{Gauss–Hermite quadrature}

\acrodef      {AGHQ}{adaptive Gauss–Hermite quadrature}

\acrodef      {GWI}{Gaussian weighted integral}
\acrodefplural{GWI}{Gaussian weighted integrals}

\acrodef      {EM}{expectation maximization}
\acrodef      {ECM}{expectation conditional maximization}

\acrodef      {GHK}{Geweke-Hajivassiliou-Keane simulator}

\acrodef      {QMC}{quasi-Monte Carlo}

\acrodef      {RQMC}{randomized quasi-Monte Carlo}

\IfFileExists{upquote.sty}{\usepackage{upquote}}{}
\begin{document}

\title{%
  Approximation Methods for Mixed Models with Probit Link Functions}

\author{
  Benjamin Christoffersen\textsuperscript{1,2,3} \and 
  Mark Clements\textsuperscript{1,3} \and 
  Hedvig Kjellström\textsuperscript{1,3} \and
  Keith Humphreys\textsuperscript{2,3} \\[.2em]
  \small\textsuperscript{1}Karolinska Institutet, MEB, Stockholm, Sweden \\[-.33em]
  \small\textsuperscript{2}KTH Royal Institute of Technology, RPL, Stockholm, Sweden \\[-.33em]
  \small\textsuperscript{3}The Swedish e-Science Research Center, Stockholm, Sweden
}

\maketitle



%


\begin{abstract}
We study approximation methods for a large class of mixed models with a
probit link function that includes mixed versions of the binomial model,
the multinomial model, and generalized survival models. 
The class of models is special because the marginal likelihood can be expressed 
as Gaussian weighted integrals or as multivariate Gaussian cumulative density 
functions. The latter approach is 
unique to the probit link function models and has been proposed for 
parameter estimation in complex, mixed effects models. However, it has not been 
investigated in which scenarios either form is preferable. Our simulations and 
data example show that neither form is preferable in 
general and give guidance on
when to approximate the cumulative density functions and when to approximate
the Gaussian weighted integrals and, in the case of the latter, which
general purpose method to use among a large list of methods. \\[.5em]
\emph{Keywords:}
Gauss–Hermite quadrature; mixed models; %
multivariate normal CDF approximation; probit link; %
randomized quasi-Monte Carlo; stochastic spherical-radial rules
\end{abstract}

\clearpage

\section{Introduction}

Estimating mixed models with moderate-to-high dimensional random effect 
terms per cluster can be computationally expensive. 
This is particularly true for some  
discrete choice models in economics, family-based 
analysis of genetic and environmental contributions with both survival times 
and categorical outcomes in biostatistics, and grouped measurements 
with censoring.
Most often, analytical 
expressions of the log marginal likelihood term for each cluster and the
derivatives with respect to the model parameters do not exist.
In this paper, we focus on a class of mixed models
which have in common that the intractable log marginal likelihood terms 
can be written in different ways, in terms of the log
of two different types of intractable integrals. The first type of 
intractable integrals is \acp{GWI} for which many general integral 
approximation methods have been suggested. 
For the second type of intractable integrals, one can show that the 
intractable integrals can be written as multivariate normal \acp{CDF}.
This is unique for this class of mixed models.

The class of models includes: mixed binomial models with a probit link 
such as those
used by \cite{Ochi84}, \cite{Pawitan04}, and \cite{Lichtenstein09};
the discrete time 
survival submodel used by \cite{Barrett15} as part of a joint model; 
mixed multinomial models with a probit 
link such as in the conditional model used by \cite{Girolami06}; 
mixed ordinal models, 
mixed \acp{GSM} with a probit link \citep{Royston02,Liu16,Liu17}; and 
a mixed version of the linear transformation model with a probit link 
\citep{Hothorn18}. The class also include heterogeneous types of data such 
as in Gaussian copulas \citep{Masarotto12,christoffersen21}.

For the above models, 
the marginal likelihood factor for each cluster is commonly written as a
\ac{GWI} when a multivariate normal distribution is assumed for the 
cluster specific random effects. Typically, the factor for each cluster is 
intractable but can be approximated using a Laplace approximation 
\citep{Lindstrom90,Wolfinger93}, \ac{GHQ}, \ac{AGHQ} \citep{Pinheiro95}, 
or some \ac{MC} method such as importance sampling.
However, an application of a standard result from the generalized 
skew-normal distribution shows that 
the marginal likelihood can also, as stated above, be written as a product 
of 
multivariate normal \acp{CDF} \citep{Azzalini05,Arnold09} or as a 
mixture of \acp{GWI} and \acp{CDF}, as has been done for a number of 
particular models \citep{Ochi84,Pawitan04,Masarotto12,Barrett15}.

We will refer to approximating the multivariate normal \acp{CDF}, rather than 
the \acp{GWI}, as the \ac{CDF} approach. 
A large number of algorithms have been developed for the multivariate normal 
\ac{CDF} \citep{Genz09} some of which are fast, even for moderate-to-high 
dimensional integrals. However, these methods are still approximations 
and, as we will show, the dimension of the \acp{CDF} may be much larger than 
the \acp{GWI} in some cases, making the \ac{CDF} approach less attractive.
Nevertheless some researchers have suggested 
the \ac{CDF} approach may generally be preferable \citep{Barrett15}.

There are two main contributions in this paper. 
Firstly, we show how to derive the formulae needed to apply the \ac{CDF} 
approach for mixed binomial models, mixed ordinal models, mixed 
multinomial models, and mixed \acp{GSM}. This is done by showing a general 
class of models which can be applied in other settings than those we show. 
Secondly, we study when the \ac{CDF}
approach is preferable and, if not, which method to use for the 
\ac{GWI} using 
simulations studies and an observational data set. In particular,
we compare the \ac{CDF} approach with approximations of the \ac{GWI} 
using \ac{GHQ}, \ac{AGHQ}, the stochastic spherical-radial rules 
shown by \cite{Genz98}, and a \ac{RQMC} method. The particular 
\ac{RQMC} method we use is Sobol sequences \citep{Bratley88,Joe03} using 
the scrambling method used by \cite{Owen98}.
Our simulation studies and application with an observational data set
show that none of the 
approximation methods are uniformly superior to all others.

The software we have written is available at 
https://github.com/boennecd/\-mixprobit. 
It is programmed \Cpp ~with our own interface to the third 
party Fortran code e.g.\ for the  stochastic spherical-radial rules 
\citep{Genz99}, with some changes to the Fortran code to allow computation 
in parallel, gradient approximations and more.
Parts of the code we have written 
may already be useful to practitioners as they are, or as components
in a procedure. 

The structure of the paper is as follows. We introduce 
the general class of mixed models in Section \ref{sec:class}.
Section \ref{sec:Aprx} contains a description of 
the general integral approximation methods we use, along with the 
approximation we use for the \ac{CDF} approach. 
Specific models are shown Section \ref{sec:models}. 
In Section \ref{sec:SimEx}, we perform simulation studies where we compare 
computation times of the approximation methods at a 
fixed precision level.
An application is provided in Section \ref{sec:application}, in which 
we show that the \ac{CDF} approximation and a \ac{MC} approximation
are fast and yield an improvement
compared to a Laplace approximation which is 
very fast but yields biased estimates as shown by others. 
We end with a discussion in Section \ref{sec:Conc}.
Alternative approximation methods and estimation methods which we do not 
cover are discussed in Appendix \ref{sec:alternatives}.

\subsection*{Notation}
We denote vectors in lower case with bold font, e.g.\ $\vec v$, and 
matrices by upper case in bold font, e.g.\ $\mat X$. All scalar functions 
will be applied element wise. That is, for $\vec x = (x_1, x_2)^\top$ then 
$f(\vec x) = (f(x_1), f(x_2))^\top$. This also applies to operators like 
`$\cdot$'. $p$ will denote a (conditional) density 
function or probability mass function. The definition is implicitly
given by the context and arguments passed to $p$. 

$\phi(x;\mu,\sigma^2)$ and $\Phi(x;\mu,\sigma^2)$ are the \ac{PDF} and \ac{CDF}, respectively, for a normal distribution with mean 
$\mu$ and standard deviation $\sigma$. We also define $\phi(x) = \phi(x; 0, 1)$
and $\Phi(x) = \Phi(x; 0, 1)$ as shorthands for the standard normal distribution. 
Further, we define %
$$\Phi^{(k)}(\vec x; \vec\mu, \mat\Sigma) = \Prob(\vec X \leq \vec x), 
\qquad \vec x \in \mathbb{R}^k$$%
where $\vec X \sim N^{(k)}(\vec\mu,\mat\Sigma)$.
$\phi^{(k)}$ will denote the multivariate normal distribution's PDF and we will 
use similar shorthands of $\phi^{(k)}$ and $\Phi^{(k)}$ for the standard case.
$\vec 0$ is a vector of zeros and $\vec 1$ is a vector of ones. The length 
of the vectors is implicit given by the context to match the vector operations. 
Similarly, $\mat I$ denotes the identity matrix.
$\diag(\cdot)$ returns the diagonal if a single square
matrix is passed as the input. $\diag(\mat{A}_1, \dots, \mat{A}_n)$ returns 
a block diagonal matrix whose block matrices are 
$\mat{A}_1, \dots, \mat{A}_n$.


\section{The Class of Mixed Models}\label{sec:class}

First, we define the general formula for the log marginal likelihood which 
the mixed models we consider have in common. All the mixed models we
use share that the outcomes can be expressed as a transformation of a latent 
multivariate normal distributed variable. 
The models have in common that the log marginal likelihood term for each 
cluster can be written as some function of the outcomes, which has an 
analytical expression, plus the log of a \ac{GWI} or multivariate normal 
\ac{CDF}. To be specific, let 
$\vec y_i = (\vec y_{i1}, \dots, \vec y_{in_i})^\top$ denote the $n_i$ 
observed outcomes for cluster $i$. Let $\vec U_i\in\mathbb{R}^K$ denote 
unobserved cluster specific 
random effects, and let $\mat\Sigma$ and $\vec\theta$ be unknown model 
parameters. Then the complete data likelihood factor for the $i$th 
group is %
\begin{equation}\label{eqn:GenericLCDF}
p(\vec u_i, \vec y_i) = c_i(\vec y_i; \vec\Sigma, \vec\theta)
   \phi^{(K)}(\vec u_i; \vec\mu_i(\vec y_i), \mat\Sigma)
   h_i(\vec y_i, \vec u_i, \vec\theta)
\end{equation}%
where $c_i$, $\vec\mu_i$, and $h_i$ have an analytical 
expression and are easy to evaluate for a given $\vec u_i$. 
It then follows that the log marginal likelihood term is %
\begin{equation}\label{eqn:GenericMargLL}
l_i(\vec\theta, \mat\Sigma) = \log c_i(\vec y_i; \vec\Sigma, \vec\theta)
  + \log\int
  \phi^{(K)}(\vec u; \vec\mu_i(\vec y_i), \mat\Sigma)
  h_i(\vec y_i, \vec u, \vec\theta)
  \der\vec u
\end{equation}%
which is the log of some function with an analytical expression, $c_i$, 
plus the log of a 
\ac{GWI}. The log marginal likelihood given $i = 1, \dots, G$ groups is %
\begin{equation}\label{eqn:MargLLSum}
\sum_{i = 1}^G l_i(\vec\theta, \mat\Sigma)
\end{equation}%
In what follows, we drop the cluster index, $i$, to ease the notation. 
We only need to derive 
the log marginal likelihood for a single group 
or the complete data likelihood for a single group in order to get the 
log marginal likelihood in Equation \eqref{eqn:MargLLSum}. Thus, we will 
e.g.\ write $\vec y$ when we implicitly mean $\vec y_i$ for some $i$.

What is particular for the models that we consider is that we can 
also write the complete data likelihood as %
\begin{equation}\label{eqn:GenericCompLCDF}
p(\vec u, \vec y) = c(\vec y; \vec\Sigma, \vec\theta)
   \phi^{(K)}(\vec u; \vec\mu(\vec y), \mat\Sigma)
   \Phi^{(k(\vec y))}
   (\vec\eta(\vec y, \vec\theta) + \mat Z(\vec y)\vec u; 
    \vec 0, \mat\Omega)
\end{equation}%
where $k(\vec y)$, $\vec\eta(\vec y, \vec\theta)$,  
$\mat Z(\vec y)$, and $\mat\Omega$
are known and may be specific to the cluster. 
Thus, the log marginal likelihood term can also be written as%
\begin{equation}\label{eqn:GenericMargLLCDF}
l(\vec\theta, \mat\Sigma) = \log c(\vec y; \vec\Sigma, \vec\theta)
  + \log\int
  \phi^{(K)}(\vec u; \vec\mu(\vec y), \mat\Sigma)
  \Phi^{(k(\vec y))}
  (\vec\eta(\vec y, \vec\theta) + \mat Z(\vec y)\vec u; 
   \vec 0, \mat\Omega)
  \der\vec u
\end{equation}%
where it is computationally advantage to use that when $k(\vec y) < K$ %
\begin{multline}
\int
  \phi^{(K)}(\vec u; \vec\mu(\vec y), \mat\Sigma)
  \Phi^{(k(\vec y))}
  (\vec\eta(\vec y, \vec\theta) + \mat Z(\vec y)\vec u; 
   \vec 0, \mat\Omega)
  \der\vec u \\
  = \int
  \phi^{(k(\vec y))}(\vec u; \mat Z(\vec y)\vec\mu(\vec y), 
  \mat Z(\vec y)\mat\Sigma\mat Z(\vec y)^\top)
  \Phi^{(k(\vec y))}
  (\vec\eta(\vec y, \vec\theta) + \vec u; 
   \vec 0, \mat\Omega)
  \der\vec u\label{eqn:simplifyGWI}
\end{multline}

We will now show that the intractable integral in the above log marginal 
likelihood term can be expressed as the log of a \ac{GWI}, as it is now,
or as the log of a multivariate normal \ac{CDF}.
The latter turns out to be advantageous as it may be preferable to 
approximate the 
\ac{CDF} rather than the \ac{GWI} in some cases, as our simulation
examples show in Section \ref{sec:SimEx}. 

\subsection{Generalization of the Skew-normal Distribution}
We will need a few results from a generalization of the multivariate skew-normal 
distribution 
in order to derive an alternative expression for 
the log marginal likelihood and the conditional density of the random effects 
given the observed outcomes. These are stated in \cite{Azzalini05,Arnold09}. 
Consider two random vectors $\vec V_1$ and $\vec V_2$ such that %
\begin{equation}\label{eqn:exGenSkew}
\begin{pmatrix}
  \vec V_1 \\ \vec V_2
\end{pmatrix} \sim 
  N^{(k_1 + k_2)}\left(
  \begin{pmatrix}
  \vec \xi_1 \\ \vec\xi_2
  \end{pmatrix}, 
  \begin{pmatrix}
  \mat\Xi_{11} & \mat\Xi_{12} \\
  \mat\Xi_{21} & \mat\Xi_{22}
  \end{pmatrix}
  \right)
\end{equation}%
where $\vec V_1\in\mathbb{R}^{k_1}$ and $\vec V_2\in\mathbb{R}^{k_2}$, 
$\vec\xi_1$ and $\vec\xi_2$ are mean vectors for $\vec V_1$ and $\vec V_2$, 
respectively,
and $\mat\Xi$ is a covariance matrix which we have decomposed into four 
parts. 
Then the joint density of $\vec V_1 = \vec v_1$ and $\vec V_2 \leq \vec v_2$ is %
\begin{align}
\phi^{(k_1)}(\vec v_1; \vec \xi_1, \mat\Xi_{11})
  \Prob\left(\vec V_2 < \vec v_2 \,\middle\vert\,
  \vec V_1 = \vec v_1
  \right)\nonumber
  \hspace{-150pt}& \\
&= \phi^{(k_1)}(\vec v_1; \vec \xi_1, \mat\Xi_{11})
  \Phi^{(k_2)}\left(
  \vec v_2
   - \mat\Xi_{21}\mat\Xi_{11}^{-1}(\vec v_1 - \vec\xi_1);
  \vec \xi_2, 
  \mat\Xi_{22} - \mat\Xi_{21}\mat\Xi_{11}^{-1}\mat\Xi_{12}
  \right)\label{eqn:skewDenMatch}
\end{align}%
and the marginal for $\Prob(\vec V_2 \leq \vec v_2)$ is %
\begin{align}
\Prob(\vec V_2 \leq \vec v_2) &= 
  \Phi^{(k_2)}(\vec v_2; \vec\xi_2, \mat\Xi_{22}) = 
  \Phi^{(k_2)}(\vec 0; \vec\xi_2 - \vec v_2, \mat\Xi_{22}) 
  \label{eqn:margSkewCondLHS}\\
&= 
  \int \phi^{(k_1)}(\vec v_1; \vec \xi_1, \mat\Xi_{11})
  \Prob\left(\vec V_2 < \vec v_2 \,\middle\vert\,
  \vec V_1 = \vec v_1
  \right)\der\vec v_1
  \label{eqn:margSkewCondRHS}
\end{align}%
Both the integral in Equation \eqref{eqn:margSkewCondLHS}, the \ac{CDF}, 
and the integral in Equation \eqref{eqn:margSkewCondRHS}, the \ac{GWI}, 
are intractable but their dimensions may differ greatly. 
Notice that the integrand in Equation \eqref{eqn:margSkewCondRHS} match the 
integrand in Equation \eqref{eqn:GenericMargLLCDF}. Thus, we can use the 
identity in Equation \eqref{eqn:margSkewCondLHS} and 
\eqref{eqn:margSkewCondRHS} to express the $K$-dimensional \ac{GWI}
in the log marginal likelihood term as a $k(\vec y)$-dimensional 
multivariate normal \ac{CDF}. Specifically, we find that %
\begin{align*}
\vec\xi_1 &= \vec\mu(\vec y) & 
  \vec\xi_2 &= \vec 0 & 
  \mat\Xi_{11} &= \mat\Sigma & 
  \mat\Xi_{21} &= -\mat Z(\vec y)\mat\Sigma \\
  \mat\Xi_{22} &= \mat\Omega + \mat Z(\vec y)\mat\Sigma\mat Z(\vec y)^\top & 
  k_1 &= K & 
  k_2 &= k(\vec y) & 
  \vec v_2 &= \vec\eta(\vec y, \vec\theta) 
    - \mat\Xi_{21}\mat\Xi_{11}^{-1}\vec\mu(\vec y)
\end{align*}%
Thus, the log marginal likelihood in Equation \eqref{eqn:GenericMargLLCDF} can 
be expressed as%
\begin{align*}
l(\vec\theta, \mat\Sigma) &= \log c(\vec y; \vec\Sigma, \vec\theta)
  + \log\int
  \phi^{(K)}(\vec u; \vec\mu(\vec y), \mat\Sigma)
  \Phi^{(k(\vec y))}
  (\vec\eta(\vec y, \vec\theta) + \mat Z(\vec y)\vec u; 
   \vec 0, \mat\Omega)
  \der\vec u \\
&=  \log c(\vec y; \vec\Sigma, \vec\theta)
  + \log\Phi^{(k(\vec y))}\left(
  \vec 0; -\vec\eta(\vec y, \vec\theta) - \mat Z(\vec y)\vec\mu(\vec y), 
  \mat\Omega + \mat Z(\vec y)\mat\Sigma\mat Z(\vec y)^\top
  \right)
\end{align*}

Importantly, numerically approximating the \ac{CDF}
of a multivariate normal distribution,
and its derivatives 
with respect to the mean and covariance matrix, have received 
a great deal of attention. 
We discuss the particular approximations we use in Section \ref{sec:Aprx}. 
It is important to be clear that the \ac{CDF} approach to fitting mixed 
models with probit links is an approximation method. We would argue 
that some authors have not been clear about this. \cite{Barrett15}, 
for example, state that 
the multivariate normal \ac{CDF} has 
a closed form solution and thus, using the \ac{CDF} in Equation 
\eqref{eqn:margSkewCondLHS} results in exact likelihood inference. 
However, the \ac{CDF} of a multivariate normal distribution 
over a hyperrectangle has no 
analytical solution in general, and as our simulation example 
in Section \ref{sec:SimEx} shows,
it seems that,
unlike what is implied by these authors, 
it is not always preferable to use the \ac{CDF} 
approach, as opposed to the \ac{GWI} approach
with efficient \Cpp ~and Fortran implementations,
or vice-versa.

The conditional density of $\vec V_1$ given $\vec V_2\leq\vec v_2$ is 
identical to the density of a generalized skew-normal distribution. 
The conditional density is given by %
\begin{align}
p(\vec v_1\mid \vec V_2 \leq \vec v_2) &=   
  \phi^{(k_1)}(\vec v_1; \vec \xi_1, \mat\Xi_{11})
  \frac
  {\Prob(\vec V_2 \leq\vec v_2 \mid \vec V_1 = \vec v_1)}
  {\Prob(\vec V_2 \leq\vec v_2)} \nonumber\\
&= \phi^{(k_1)}(\vec v_1; \vec \xi_1, \mat\Xi_{11})
  \frac
  {\Phi^{(k_2)}\left(
  \vec v_2
   - \mat\Xi_{21}\mat\Xi_{11}^{-1}(\vec v_1 - \vec\xi_1);
  \vec \xi_2, 
  \mat\Xi_{22} - \mat\Xi_{21}\mat\Xi_{11}^{-1}\mat\Xi_{12}
  \right)}
  {\Phi^{(k_2)}\left(
  \vec v_2; \vec \xi_2, \mat\Xi_{22}
  \right)}\label{eqn:skewCond}
\end{align}%
which is equivalent to the conditional density of the random effects given the 
outcomes for the class of mixed models we work with. 
A random variable follows a generalized skew-normal distribution if its 
density is given by \eqref{eqn:skewCond} where $\vec v_2$, 
$\vec\xi_1$, $\vec\xi_2$, 
$\mat\Xi_{11}$, $\mat\Xi_{12} = \mat\Xi_{21}^\top$, and $\mat \Xi_{22}$ are 
parameters in the family. 
We now turn to the approximation method we use in the \ac{CDF} approach and 
the general integral approximation methods we consider to 
approximate the \ac{GWI}.


\section{Approximation Methods for Mixed Effects Models with Probit Link 
Functions}\label{sec:Aprx}
In this section we describe the approximation methods that we 
use. We start with the \ac{CDF} approach where 
we approximate the \ac{CDF} rather then the \ac{GWI}.
As mentioned previously, approximating the multivariate normal \ac{CDF} has 
received a great deal of attention.
The approach of \cite{Genz92}, 
or the similarly and seemingly independently developed \ac{GHK} 
used by \cite{Hajivassiliou96}, is fast and quite precise. Moreover, the 
implementation in the \texttt{mvtnorm} package \citep{Genz09, mvtnorm} in R 
\citep{R19} uses 
randomized Korobov rules 
\citep{Niederreiter1972,Keast73, Cranley76}
which yields a huge reduction in computation time compared to the \ac{MC}
implementation of the method described in \cite{Genz92}.
Thus, the \ac{CDF} approximation we are using is also a \ac{RQMC} method 
(Section \ref{subsec:RQMC}). We note that we have
adapted the Fortran code for the \ac{CDF} approximation to also provide gradient
approximations with respect to the mean and covariance matrix. This resulted in a 
more than four-fold reduction in the computation time in the example 
in Section \ref{sec:application}. 
See \cite{Hajivassiliou96} for a thorough comparison of 
methods to approximate the multivariate normal \ac{CDF} and its derivatives.

The method suggested by \cite{Genz92} works as follows. Let 
$\mat\Sigma = \mat S\mat S^\top$ where $\mat S$
is the Cholesky decomposition of $\mat\Sigma$. Then %
\begin{align*}
\Phi^{(k)}(\vec l, \vec b; \vec\mu, \mat\Sigma) &= 
  \int_{l_1}^{b_1}\cdots\int_{l_k}^{b_k} 
  \phi^{(k)}(\vec v;\vec\mu, \mat\Sigma)
  \der v_1\cdots \der v_k \\
  &=
  \int_{\tilde l_1}^{\tilde b_1}\int_{\tilde l_2(v_1)}^{\tilde b_2(v_1)}
  \cdots\int_{\tilde l_k(v_1, \dots, v_{k-1})}^{\tilde b_k(v_1, \dots, v_{k-1})}
  \phi^{(k)}(\vec v)\der v_1\cdots \der v_k \\
  &= 
  \int_{\tilde l_1}^{\tilde b_1}\phi(v_1)
  \int_{\tilde l_2(v_1)}^{\tilde b_2(v_1)}\phi(v_2)
  \cdots\int_{\tilde l_k(v_1, \dots, v_{k-1})}^{\tilde b_k(v_1, \dots, v_{k-1})}
  \phi(v_k)
  \der v_1\cdots \der v_k
\end{align*}%
where %
\begin{align*}
\tilde l_1 &= s_{11}^{-1}(l_1 - \mu_1) &
  \tilde l_k(v_1, \dots, v_{k-1}) &= s_{kk}^{-1}
  \left(l_k  - \mu_k - \sum_{i=1}^{k-1}s_{ki}v_i
  \right) \\
 \tilde b_1 &= s_{11}^{-1}(u_1 - \mu_1) &
  \tilde b_k(v_1, \dots, v_{k-1}) &= s_{kk}^{-1}
  \left(b_k  - \mu_k - \sum_{i=1}^{k-1}s_{ki}v_i
  \right)
\end{align*}%
This suggests the following \ac{MC} estimator: sample $\hat v_1$ from a 
truncated normal distribution truncated at $\tilde l_1$ and $\tilde b_1$. 
Then sample $\hat v_2$ from truncated normal distribution truncated at 
$\tilde l_2(\hat v_1)$ and $\tilde b_2(\hat v_1)$. Repeat sampling till 
reaching $\hat v_{k - 1}$. Then evaluate %
\begin{equation}\label{eqn:GenzMCEst}
  (\Phi(\tilde b_1) - \Phi(\tilde l_1))
  \prod_{i = 2}^k\left(
  \Phi(\tilde b_i(\hat v_1, \dots, \hat v_{i-1})) - 
  \Phi(\tilde l_i(\hat v_1, \dots, \hat v_{i-1}))
  \right)
\end{equation}%
Repeat for a new sample and average the results till the estimated \ac{MC}
error is below a prespecified threshold. A greedy variable reordering 
either to increase the expected width of the inner most 
integrals or to reduce the variance of Equation \eqref{eqn:GenzMCEst} can 
yield a substantial reduction in variance at a fixed number of samples
in some cases \citep[Section 4.1]{Genz09}. 

Because of the aforementioned method and its precision, approximating the 
$k_2$-dimension\-al \ac{CDF} might be attractive even 
when $k_2$ is larger 
than $k_1$. 
However, it should be noticed that if $k_1$ is much smaller than 
$k_2$, then 
approximating the \ac{GWI} in  
Equation \eqref{eqn:margSkewCondRHS} may be more attractive 
than approximating the \ac{CDF} in Equation \eqref{eqn:margSkewCondLHS},
as we will show in Section \ref{sec:SimEx}. 
For the models described in this paper, $k_1 = K$ and $k_2 = k(\vec y)$. 
Only the latter is model specific. 

\subsection{Approximating the Gaussian Weighted Integral}\label{subsec:AprxGWI}
We now focus on approximating the \ac{GWI}. That is, integrals of the form %
\begin{align}
g(\vec u) &= 
  \phi^{(k_1)}(\vec u; \vec \xi_1, \mat\Xi_{11})
  \Prob\left(\vec V_2 < \vec v_2 \,\middle\vert\,
  \vec V_1 = \vec u 
  \right) 
  \label{eqn:Integrand}\\
L &= \int g(\vec u)\der \vec u
  \label{eqn:GeneralMargL}
\end{align}%
where we let $g(\vec u)$ denote the integrand and
$\Prob(\vec V_2 < \vec v_2 \,\vert\,\vec V_1 = \vec u )$ is a 
$k_2$-dimensional \ac{CDF} as in Equation \eqref{eqn:skewDenMatch}. 

It is worth remarking that the structure of the conditional \ac{CDF},
$\Prob(\vec V_2 < \vec v_2 \,\vert\,\vec V_1 = \vec u )$, should 
be used for a given model; for example, this expression can be evaluated 
as products of 
powers of $n$ 
\acp{CDF} of the standard normal distribution in the mixed 
binomial model we cover later. This is much more 
attractive than approximating the $k_2$-dimensional conditional 
\ac{CDF} in the integrand
as a general \ac{CDF}. Moreover, 
special structure may be exploited in certain cases as done by 
\cite{Ochi84}.
Because of the special structure, approximating the 
\ac{GWI} may be faster than approximating the \ac{CDF} at a fixed precision.
Lastly, Equation \eqref{eqn:simplifyGWI} should be used 
when $k(\vec y) < K$. 

\subsubsection{Laplace Approximation}\label{subsec:Laplace}
One way to approximate the \ac{GWI} is to use a Laplace approximation. 
This begins with computing a mode%
$$
\widehat{\vec u} = \argmax_{\vec u} \log g(\vec u)
$$%
and the Hessian %
$$
\widehat{\mat H} = \left. 
  \frac{\partial^2}{\partial \vec u \partial \vec u^\top} 
  \log g (\vec u) \right\vert_{\vec u = \widehat{\vec u}}
$$%
The Laplace approximation is then %
$$
\int g(\vec u)\der \vec u \approx
  \frac {(2\pi)^{K/2}}{
  \lvert- \widehat{\mat H}\rvert^{1 / 2}}
  g(\widehat{\vec u})
$$

\cite{Lindstrom90} and \cite{Wolfinger93} show a version of the Laplace 
approximation in which fixed effects are also estimated. The Laplace 
approximation is a very fast means of approximating the log marginal 
likelihood, but may have a substantial bias in some cases 
\citep{Raudenbush00,Pinheiro06,Joe08}. 
The computational complexity is $\bigO{K^3 + nK}$ if we 
assume that the $K$-dimensional mode can be estimated at rate, at or 
below, $\bigO{K^3 + nK}$.

\subsubsection{Simple Monte Carlo Estimator}\label{sec:SimpleMC}
For the simulations, 
our ground truth estimator will be a simple \ac{MC} estimator 
where we use a large number of samples. A simple procedure for 
approximating the marginal likelihood in Equation \eqref{eqn:GeneralMargL} 
is to
sample $\vec u \sim N^{(k_1)}(\vec \xi_1, \mat\Xi_{11})$, 
evaluate $\Prob(\vec V_2 < \vec v_2 \,\vert\,\vec V_1 = \vec u )$, 
and average the outcomes. An adaptive and yet still simple alternative,
is to compute $\widehat{\vec u}$ and $\widehat{\mat H}$ as we do with the
Laplace approximation. Then letting %
\begin{equation}\label{eqn:AdaIntegrand}
\widehat g(\vec u) = 
  \phi^{(k_1)}
  \left(\vec u; \widehat{\vec u}, (-\widehat{\mat H})^{-1}\right)
  \frac
  {\phi^{(k_1)}\left(\vec u; \vec \xi_1, \mat\Xi_{11}\right)}
  {\phi^{(k_1)}\left(\vec u; \widehat{\vec u}, (-\widehat{\mat H})^{-1}
   \right)}
  \Prob\left(\vec V_2 < \vec v_2 \,\middle\vert\,
  \vec V_1 = \vec u 
  \right),
\end{equation}%
we perform importance sampling by sampling 
$\vec u \sim N^{(k_1)}(\widehat{\vec u}, (-\widehat{\mat H})^{-1})$,
evaluating the rest of the factors in the integrand $\widehat g$, 
and averaging the 
outcomes.
The additional computation cost of finding the mode, 
computing the Hessian,
and the overhead in the integrand is typically much smaller compared to 
the reduction in the variance of the estimator. 

Our \Cpp ~implementation also uses three simple antithetic variables 
for each sampled $\vec u$ to reduce the variance of the estimator.
These are the so-called location and scale balanced variables in the 
terminology of \cite{Durbin97}.
A major advantage of this \ac{MC} estimator and all the other \ac{MC}
estimators that we discuss is that a standard error of the estimator can easily be computed 
at any iteration of the approximation. Thus, a stopping criteria can be 
implemented, as is done in the \ac{CDF} approximation in the 
\texttt{mvtnorm} package.

\subsubsection{Alternative Monte Carlo Estimator}\label{sec:GWIGenz}
The variance of the importance sample estimator at a fixed computing budget 
may still be unacceptable. Thus, we will use the \texttt{RANRTH} Fortran 
subroutine developed by \cite{Genz99} as an alternative \ac{MC} estimator.
This subroutine uses stochastic 
spherical-radial rules to approximate \acp{GWI} like 
Equation \eqref{eqn:GeneralMargL}. The degree one rule is identical to using 
the location balanced variables that we use with our importance sampler. 
\cite{Genz99}
show higher degree rules. We will use the 
five degree rule.
We changed the implemented Householder reflection to use the algorithm
shown in
\citet[Section 5.1]{Golub13}. The previous algorithm is numerically unstable 
and caused problems in our simulations.

Let $s$ be the number of samples that is 
used. Then the complexity is $\mathcal O(K^3 + s(nK + \allowbreak K^2 ))$ 
with the rule that we use. 
As we show later, using an adaptive approach as suggested by
\cite{Genz98}, similar to what we do with the 
importance sampler in Section \ref{sec:SimpleMC}, 
yields much more precise results at a fixed 
computing budget.
Our main reason to include the method suggested by \cite{Genz99} is not to 
argue to use this particular \ac{MC} estimator but to show one \ac{MC} 
estimator of the \ac{GWI} which may scale better in the number of 
random effect terms, $K$, than the \ac{GHQ} and 
\ac{AGHQ} approaches which we cover in Section \ref{subsec:GWIGHQ}. 

\subsubsection{Randomized Quasi-Monte Carlo}\label{subsec:RQMC}

\begin{knitrout}
\definecolor{shadecolor}{rgb}{0.969, 0.969, 0.969}\color{fgcolor}\begin{figure}

{\centering \includegraphics[width=\maxwidth]{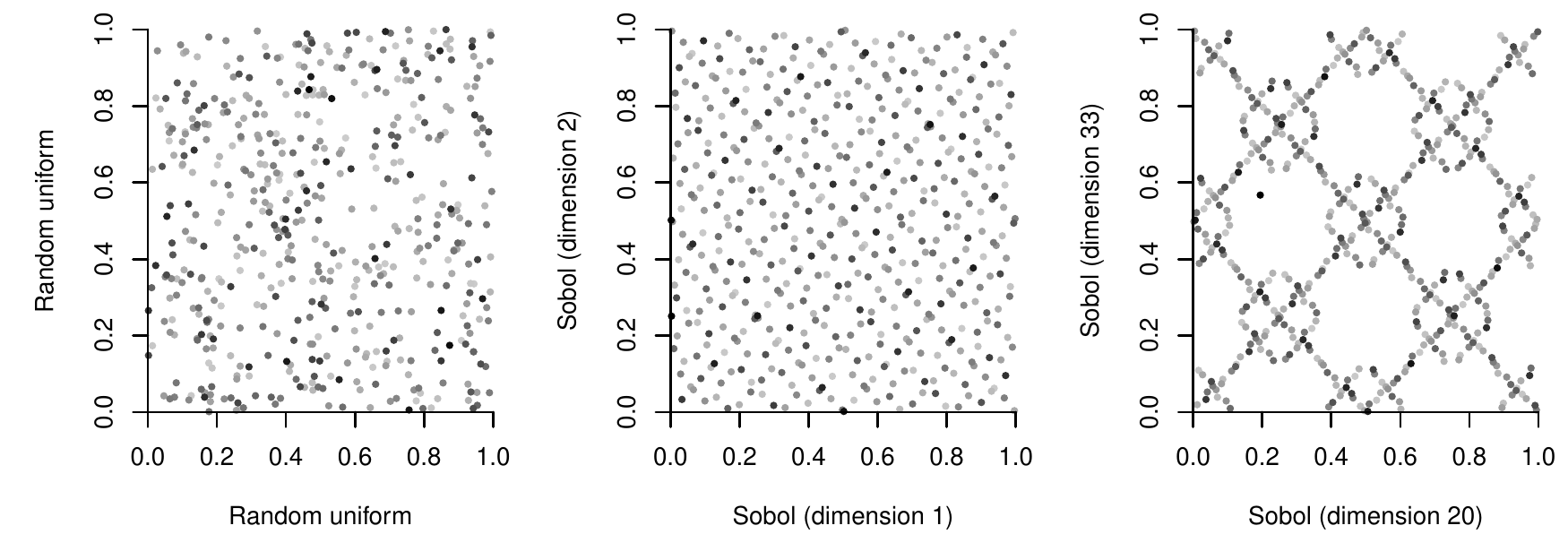} 

}

\caption[The illustration compares \acs{MC} with \acs{QMC}]{The illustration compares \acs{MC} with \acs{QMC}. All three figures show 512 points. The left figure shows points sampled from the uniform distribution over $[0,1)^2$. The figure in the middle shows the first two dimensions from a Sobol sequence and the right figure shows the 20th and 33th dimensions. Darker colors come later in the sequences to show that points from the Sobol sequence tends to be further apart if they are closer to each other in the sequence.}\label{fig:showQMC}
\end{figure}

\end{knitrout}

\ac{MC} methods have a $\bigO{s^{-1/2}}$ convergence rate
of the estimator where $s$ is the number of samples. This can make it 
very expensive to achieve a given accuracy. An 
alternative is to consider \ac{QMC} methods. 
\ac{QMC} methods have an error bound 
which is $\bigO{s^{-1}(\log s)^K}$ but the convergence rate is often lower 
in practical applications. \ac{QMC} methods achieve a better convergence 
rate by selecting a sequence of points in $[0, 1)^K$ which is more 
uniformly spread than a random sequence. We illustrate this in Figure 
\ref{fig:showQMC} where we sample uniformly from $[0,1)^K$ and compare the 
sample with the Sobol sequence. The uniformity is measured by  
a sequence's discrepancy which can be used to bound the error with 
the Koksma–Hlawka inequality \citep{caflisch98}. 

We use \ac{RQMC} and particularly scrambled Sobol sequences with 
the scrambling method used by \cite{Owen98}. The Fortran code is from 
the \texttt{randtoolbox} package 
\citep{randtoolbox} and is based on the code implemented by \cite{Bratley88} 
and \cite{Joe03}. One advantage of \ac{RQMC} is that we can get a standard 
error for the estimate. 
This has allowed us to implement a stopping criteria similar to 
that available with the \ac{MC} methods we use. We have also used an 
adaptive procedure with \ac{RQMC}. We transform the $[0,1)^K$ 
samples, first by applying the 
inverse standard normal \ac{CDF} elementwise, as we do with our \ac{MC}
methods. However, we then multiply by $\mat Q\mat\Lambda^{1/2}$, where 
$(-\mat H)^{-1} = \mat Q\mat \Lambda\mat Q^\top$ is the eigendecomposition 
rather than a Cholesky decomposition of $(-\mat H)^{-1}$. The motivation is 
that the low-dimensional projections obtained by grouping coordinates  of 
the Sobol sequence that we use, do not have holes for a small number of samples 
with the leading dimensions, although with the later dimensions they do, 
as shown in Figure \ref{fig:showQMC} \citep{Morokoff94}. Thus, we ensure 
that the first dimensions have roughly the highest variation of the 
integrand by using $\mat Q\mat\Lambda^{1/2}$ and these do not have holes. 

\subsubsection{Gauss–Hermite Quadrature}\label{subsec:GWIGHQ}
In some cases, it may be preferable to use some deterministic approximation,
such as using some type 
of quadrature, for the \ac{GWI} in Equation \eqref{eqn:GeneralMargL}, when 
$K$ is small. In particular, \ac{GHQ} and \ac{AGHQ} are often used 
\citep{Liu94,Pinheiro95}. An application of \ac{GHQ} involves 
change-of-variables such that the integrand is %
$$
\phi^{(k_1)}(\vec u; \vec 0, 2\mat I)
  \Prob\left(\vec V_2 < \vec v_2 \,\middle\vert\,
  \vec V_1 = \vec \xi_1 + \sqrt 2\mat\Xi^{1/2}\vec u 
  \right), 
$$%
where $\mat\Xi^{1/2}$ is a Cholesky decomposition of $\mat\Xi$ such that
$\mat\Xi = \mat\Xi^{1/2}\mat\Xi^{\top/2}$. Direct application of 
\ac{GHQ} is then applicable to each of the $K$ random effect terms.
This implies $b$ values 
for each of the $K$ random effect terms, resulting in $b^K$ so-called nodes 
at which the integrand is evaluated. Thus, the  
computational complexity is $\bigO{(K^2 + nK)b^K}$. 

Using an adaptive procedure such as 
suggested by \cite{Pinheiro95} involves a similar application of \ac{GHQ} 
but for the integral whose integrand is shown in Equation 
\eqref{eqn:AdaIntegrand}.
This may require substantially fewer nodes, $b^K$, 
and result in much 
faster computation times at a fixed precision \citep{Pinheiro95,Hesketh02}, 
as is the case in the 
simulation example we provide in Section \ref{sec:SimEx}. 
While both \ac{GHQ} and \ac{AGHQ} 
may be fast for 
few random effect terms, i.e.\ small $K$, they will be expensive
even for few values for each random effect term, $b$, 
when there are many random effect terms.


\section{Models}\label{sec:models}
In this section we show some mixed effects models which are within the class 
of models introduced in Section \ref{sec:class}. The formulae are provided 
to apply the \ac{CDF} approach.


\subsection{Mixed Probit Model}\label{subsec:mixProb}
The first model we consider is the mixed probit model for binomial 
outcomes. 
Assume that we observe counts $\vec Y \in \{0, \dots, m\}^n$ for all individuals  
$i = 1, \dots, n$ and observe a fixed 
effect design matrix $\mat X = (\vec x_1, \dots, \vec x_n)^\top$ and 
random effect design matrix $\mat Z = (\vec z_1, \dots, \vec z_n)^\top$. 
$\vec Y$ can be generalized to non-equal counts 
such that $\vec Y \in \{0, \dots, m_1\}\times\cdots\times\{0, \dots, m_n\}$.
Moreover, 
we assume that there is an unobserved random effect $\vec U\in\mathbb{R}^K$. 
The elements of $\vec Y$ are assumed 
to be independent conditional on the random effect. 
Let $\vec\beta$ be a coefficient vector and
let $\text{Bin}$ 
denote the binomial distribution such that if $V \sim \text{Bin}(q, k)$ then %
$$
p(v) = \begin{pmatrix}k \\ v\end{pmatrix} q^v(1 - q)^{k - v}
$$%
We assume that the conditional distribution of each element of $\vec Y$ is %
$$Y_i \mid \vec U = \vec u \sim \text{Bin}(
  \Phi(\vec x_i^\top\vec\beta + \vec z_i^\top\vec u), m)$$%
That is, we are considering a \ac{GLMM} with a probit link function.
Further, we assume that the random effects follow a 
multivariate normal distribution such that %
$$\vec U \sim N^{(K)}(\vec 0, \mat \Sigma)$$%
The complete data likelihood is therefore %
\begin{align}
p(\vec u, \vec y) &= \phi^{(K)}(\vec u;\vec 0, \mat\Sigma)
  \prod_{i = 1}^n \begin{pmatrix}m \\ y_i\end{pmatrix}
  \Phi(\vec x_i^\top\vec\beta + \vec z_i^\top\vec u)^{y_i}
  (1 - \Phi(\vec x_i^\top\vec\beta + \vec z_i^\top\vec u))^{m - y_i}\nonumber \\
&= c(\vec y)  \phi^{(K)}(\vec u;\vec 0, \mat\Sigma)
  \prod_{i = 1}^n
  \Phi(\vec x_i^\top\vec\beta + \vec z_i^\top\vec u)^{y_i}
  \Phi(-\vec x_i^\top\vec\beta - \vec z_i^\top\vec u)^{m - y_i}
  \label{eqn:compL},
\end{align}%
where
$$
c(\vec y) = \prod_{i = 1}^n \begin{pmatrix}m \\ y_i\end{pmatrix}
$$%
and the log marginal likelihood is %
\begin{equation}\label{eqn:margLL}
l(\vec\beta,\mat\Sigma) = 
  \log p(\vec y) = \log\int p(\vec u, \vec y) \der\vec u, 
\end{equation}%
which is needed for model estimation in a frequentist analysis.

We will define a few augmented matrices in order to show a form of 
the complete data likelihood shown in Equation \eqref{eqn:compL} 
which is similar to Equation \eqref{eqn:GenericCompLCDF}. Let %
\begin{align}
\widetilde{\mat X}_i &= (\underbrace{\vec x_i, \dots, \vec x_i}_{
  y_i\text{ times}}, \underbrace{-\vec x_i, \dots, -\vec x_i}_{
  m - y_i\text{ times}})^\top \nonumber\\
  & = (\underbrace{1, \dots, 1}_{y_i\text{ times}}, 
       \underbrace{-1, \cdots, -1}_{m - y_i\text{ times}})^\top
       \vec x_i^\top = \vec j_i \vec x_i^\top \nonumber \\
\widetilde{\mat X} &= (\widetilde{\mat X}_1^\top, \dots, \widetilde{\mat X}_n^\top)^\top
  = \begin{pmatrix}
    \vec j_1 & \vec 0  & \cdots & \vec 0 \\
    \vec 0 & \vec j_2 & \ddots & \vec \vdots \\
    \vdots & \ddots & \ddots  & \vec 0 \\
    \vec 0 & \cdots & \vec 0 &  \vec j_n
  \end{pmatrix}\mat X = \mat J \mat X\label{eqn:XAug}
\end{align}%
We define $\widetilde{\mat Z}_i$ and $\widetilde{\mat Z}$, similarly. 
We can now rewrite the complete data likelihood as %
\begin{align}
p(\vec u, \vec y) &= c(\vec y) \phi^{(K)}(\vec u;\vec 0, \mat\Sigma)
  \prod_{i = 1}^n
  \Phi^{(m)}(\widetilde{\mat X}_i\vec\beta + \widetilde{\mat Z}_i\vec u)
  \nonumber\\
&= c(\vec y) \phi^{(K)}(\vec u;\vec 0, \mat\Sigma)
  \Phi^{(nm)}(\widetilde{\mat X}\vec\beta + \widetilde{\mat Z}\vec u)
  \label{eqn:compLAug}
\end{align}%
and the log marginal likelihood as %
\begin{align}
l(\vec\beta, \mat\Sigma) &= \log\int p(\vec u, \vec y)\der \vec u
  \nonumber \\
  &=\log c(\vec y) + \log\int 
  \phi^{(K)}(\vec u;\vec 0, \mat\Sigma)
  \Phi^{(nm)}(\widetilde{\mat X}\vec\beta + \widetilde{\mat Z}\vec u)
  \der \vec u
  \label{eqn:compLLAug}
\end{align}%
In order to find the formula needed to apply the \ac{CDF} approach, it 
follows from Equation \eqref{eqn:compLLAug} that %
\begin{align*}
\vec\xi_1 &= \vec 0 & 
  \vec\xi_2 &= \vec 0 & 
  \mat\Xi_{11} &= \mat\Sigma & 
  \mat\Xi_{21} &= -\widetilde{\mat Z}\mat\Sigma \\
  \mat\Xi_{22} &= \mat I + \widetilde{\mat Z}\mat\Sigma\widetilde{\mat Z}^\top & 
  k_1 &= K & 
  k_2 &= nm & 
  \vec v_2 &= \widetilde{\mat X}\vec\beta
\end{align*}%
Notice that $k_2 = n$ in the binary case, $m = 1$, implying that 
the \ac{CDF} approach seems advantageous particularly in the binary case. 
As an example, \cite{Pawitan04} and \cite{Lichtenstein09} use 
the \ac{CDF} approach to estimate 
a model for binary traits with family data. In this setting, the \ac{CDF} 
approach is very attractive as $K = n$ such that $k_2 = k_1$.

The random effect vector, $\vec U$, contains the random effect for a given 
cluster where all observations are only assumed independent given this 
unobserved variable. Thus, $K$ is the number of random effect terms that we 
have to integrate out. This can be defined both in a 
crossed random effects setup and
in a nested random effect setup. As an example of nested random effects, 
let each cluster be a school with six classes and 
suppose that we include both a school specific random effect 
and a class specific random effect.  
Then $K=7$, since there is a total of 7 random effect terms. We provide an 
application with a crossed random effects setup in Section 
\ref{sec:application} with $K = 20$.


\subsection{Multinomial Probit Regression}\label{subsec:Multinomial}
The second model that we consider is a multinomial probit model in which
each observed outcome, $Y_i$, can fall into one of $c$ categories. Let 
$\vec Y \in \{1,\dots,c\}^n$ denote a vector of observed outcomes,
$\mat Z_i = (\vec z_{i1}, \dots, \vec z_{ic})^\top \in \mathbb{R}^{c\times K}$ 
be a matrix of known random effect covariates for each of the $c$ 
categories for individual $i$, 
$\vec x_i$ be a vector of known fixed effect covariates for individual $i$,
$\mat B = (\vec \beta_1, \dots, \vec\beta_c)^\top$ be a matrix of 
fixed effect coefficients, and %
$$\vec A_i \mid \vec U = \vec u \sim N(\mat B\vec x_i + \mat Z_i \vec u, \mat I)$$%
be the vector of latent variables for individual $i$ such that %
$$Y_i = k \Leftrightarrow \forall k \neq k':\, A_{ik} > A_{ik'}, 
  \qquad k,k'\in\{1,\dots,c\}.$$

Let $\mat B_{(-k)} = (\vec\beta_1, \dots, \vec\beta_{k-1}, 
\vec\beta_{k + 1}, \dots, \vec\beta_c)$ be the coefficient matrix without 
the $k$th
row and similarly define $\mat Z_{i(-k)}$. 
It then follows that the conditional probability of $Y_i$ falling into
the $k$th category, given the random effect, $\vec U$,  is %
\begin{align}
\mathcal{C}_{ik} &= \left\{
  \vec A_i:\,\forall k' \neq k: A_{ik} > A_{ik'}
  \right\} 
  \nonumber\\
p(Y_i = k \mid \vec U = \vec u) &= 
  \int_{\mathcal{C}_{ik}} \phi^{(c)}
  (\vec a; \mat B\vec x_i + \mat Z_i \vec u, \mat I) 
  \der \vec a 
  \nonumber\\
&= \int
  \phi(a_k; \vec\beta_k^\top\vec x_i + \vec z_{ik}^\top\vec u, 1)
  \left(
  \prod_{k' \neq k} \int_{-\infty}^{a_k} 
  \phi(a_{k'}; \vec\beta_{k'}^\top\vec x_i + \vec z_{ik'}^\top\vec u, 1)
  \der a_{k'}\right)\der a_k  
  \nonumber\\
&= \int
  \phi(a_k; \vec\beta_k^\top\vec x_i + \vec z_{ik}^\top\vec u, 1)
  \Phi^{(c - 1)}(\vec 1 a_k - \mat B_{(-k)}\vec x_i - \mat Z_{i(-k)}\vec u)  
  \der a_k 
  \nonumber\\
&= \int
  \phi(a_k)
  \Phi^{(c - 1)}(\vec 1a_k 
  + (\vec 1\vec\beta_k^\top - \mat B_{(-k)})\vec x_i + 
  (\vec 1\vec z_{ik}^\top - \mat Z_{i(-k)})\vec u)  
  \der a_k
  \nonumber\\
&= \Phi^{(c - 1)}( 
  \underbrace{(\vec 1\vec\beta_k^\top - \mat B_{(-k)})}_{
  \widetilde{\mat B}_k}\vec x_i + 
  \underbrace{(\vec 1\vec z_{ik}^\top - \mat Z_{i(-k)})}_{
  \widetilde{\mat Z}_{ik}}\vec u;
  \vec 0, \mat I + \vec 1\vec 1^\top)
  \label{eqn:MultIntegrand}
\end{align}%
where the last equality follows from the identity in Equation 
\eqref{eqn:margSkewCondRHS}. 
See \cite{McFadden84}
for a similar derivation. Thus, the complete data 
likelihood is %
\begin{equation}\label{eqn:multInter}
p(\vec u, \vec y) = \phi^{(K)}(\vec u; \vec 0, \mat\Sigma)
  \prod_{i = 1}^n
  \Phi^{(c - 1)}( 
  \widetilde{\mat B}_{y_i}\vec x_i + 
  \widetilde{\mat Z}_{iy_i}\vec u;
  \vec 0, \mat I + \vec 1\vec 1^\top) 
\end{equation}

To get Equation \eqref{eqn:multInter} into a form like in Equation 
\eqref{eqn:GenericCompLCDF}, let
$\widetilde{\mat B} = 
    \diag(\widetilde{\mat B}_{y_1}, \dots, \widetilde{\mat B}_{y_n})$,
$\vec x = (\vec x_1^\top, \dots, \vec x_n^\top)^\top$, and
$\widetilde{\mat Z} = (\widetilde{\mat Z}_{1y_1}^\top 
  \dots, \widetilde{\mat Z}_{ny_n}^\top)^\top$. Then the complete data 
likelihood can be written as 
\begin{equation}\label{eqn:compLAugMult}
p(\vec u, \vec y) = \phi^{(K)}(\vec u; \vec 0, \mat\Sigma)
  \Phi^{(n(c - 1))}( 
  \widetilde{\mat B}\vec x + 
  \widetilde{\mat Z}\vec u;
  \vec 0, \diag(\underbrace{
  \mat I + \vec 1\vec 1^\top, \dots, \mat I + \vec 1\vec 1^\top}_{%
  n\text{ times}})). 
\end{equation} %
Thus, we find that %
\begin{align*}
\vec\xi_1 &= \vec 0 & 
  \vec\xi_2 &= \vec 0 & 
  \mat\Xi_{11} &= \mat\Sigma & 
  \mat\Xi_{21} &= -\widetilde{\mat Z}\mat\Sigma \\
  \mat\Xi_{22} &= \mat\Omega + \widetilde{\mat Z}\mat\Sigma\widetilde{\mat Z}^\top & 
  k_1 &= K & 
  k_2 &= n(c - 1) & 
  \vec v_2 &= \widetilde{\mat B}\vec x
\end{align*}%
with %
$$
\mat\Omega = \diag(\underbrace{
  \mat I + \vec 1\vec 1^\top, \dots, \mat I + \vec 1\vec 1^\top}_{%
  n\text{ times}}),
$$

Notice that the \ac{CDF} in Equation \eqref{eqn:compLAugMult} is of 
dimension $n(c-1)$ whereas the dimension of the \ac{CDF} in Equation 
\eqref{eqn:compLAug} is $nm$. As we will show later, the dimension of the CDF 
is important in terms of which approximation becomes
preferable.

\subsubsection{Ordered Case}\label{subsec:MultOrd}
Assume instead that the $c$ categories are ordered and let 
$\vec \gamma = (\gamma_0, \dots, \gamma_c)^\top = 
  (-\infty,\gamma_1,\dots, \allowbreak \gamma_{c-1}, \allowbreak \infty)^\top$
be an unknown vector of bin boundary parameters. Let $A_i$ be an unknown 
latent variable for the $i$th individual and let%
\begin{align*} 
A_i \mid \vec U = \vec u &\sim 
  N(\vec x_{i}^\top\vec\beta + \vec z_i^\top \vec u, 1) \\
Y_i & = k \Leftrightarrow 
  \gamma_{k - 1} < A_i \leq \gamma_k.
\end{align*}%
Then the complete data likelihood is %
\begin{equation}\label{eqn:MultOrdComp}
p(\vec u, \vec y) = 
  \phi^{(K)}(\vec u; \vec 0, \mat\Sigma)\prod_{i = 1}^n
  (\Phi(\gamma_{y_i} - \vec x_{i}^\top\vec\beta - \vec z_i^\top \vec u)
   - \Phi(\gamma_{y_i-1} - \vec x_{i}^\top\vec\beta - \vec z_i^\top \vec u))
\end{equation}

However, we cannot rewrite Equation \eqref{eqn:MultOrdComp} into an 
expression like 
in Equation \eqref{eqn:GenericCompLCDF}. Though, define %
$$
\Phi^{(k)}(\vec l, \vec b; \vec\mu, \mat\Sigma) = 
  \Prob(\vec l < \vec X \leq \vec b), 
  \qquad \vec X\sim N^{(k)}(\vec\mu, \mat\Sigma),
$$%
$g(k) = \gamma_k$, $\mat X = (\vec x_1, \dots, \vec x_n)^\top$, and 
$\mat Z = (\vec z_1, \dots, \vec z_n)^\top$. Then the complete data 
likelihood can be written as %
$$
p(\vec u, \vec y) =
  \phi^{(K)}(\vec u; \vec 0, \mat\Sigma)
  \Phi^{(n)}(
  g(\vec y - \vec 1) - \mat X\vec\beta - \mat Z\vec u, 
  g(\vec y         ) - \mat X\vec\beta - \mat Z\vec u)
$$

This turns out also to give an alternative expression of the log marginal 
likelihood which is the log of a multivariate normal \ac{CDF}. To see this, 
let %
\begin{multline}
\Prob\left(\vec l < \vec V_2 \leq \vec b \,\middle\vert\,
  \vec V_1 = \vec v_1
  \right) \\ 
  =\Phi^{(k_2)}\left(
  \vec l - \mat\Xi_{21}\mat\Xi_{11}^{-1}(\vec v_1 - \vec\xi_1), 
  \vec b - \mat\Xi_{21}\mat\Xi_{11}^{-1}(\vec v_1 - \vec\xi_1);
  \vec \xi_2, 
  \mat\Xi_{22} - \mat\Xi_{21}\mat\Xi_{11}^{-1}\mat\Xi_{12}
  \right)
\end{multline}%
to show that %
\begin{align*}
\Prob\left(\vec l < \vec V_2 \leq \vec b\right) &= 
  \Phi^{(k_2)}(\vec l, \vec b; \vec\xi_2, \mat\Xi_{22}) \\
  &= \int \phi^{(k_1)}(\vec v_1;\vec\xi_1, \mat\Xi_{11})
  \Prob\left(\vec l < \vec V_2 \leq \vec b \,\middle\vert\,
  \vec V_1 = \vec v_1
  \right)\der\vec v_1.
\end{align*}%
This shows that we also have a choice between a \ac{CDF} and a \ac{GWI} for 
the mixed ordered multinomial model by setting %
\begin{align*}
\vec\xi_1 &= \vec 0 & 
  \vec\xi_2 &= \vec 0 & 
  \mat\Xi_{11} &= \mat\Sigma & 
  \mat\Xi_{21} &= \mat Z\mat\Sigma \\
  \mat\Xi_{22} &= \mat I + \mat Z\mat\Sigma\mat Z^\top & 
  k_1 &= K & 
  k_2 &= n &
  \vec l &= g(\vec y - \vec 1) - \mat X\vec\beta \\
  \vec b &= g(\vec y)          - \mat X\vec\beta.
\end{align*}

Lastly, if we add 
an identification constraint of $\gamma_1 = 0$, it reduces to the mixed
binomial model in the binary case, $m = 1$, if $c = 2$. Moreover, $k_2 = n$
like in the mixed binomial model in the binary case and, importantly, this 
is independent of the number of categories, $c$.


\subsection{Generalized Survival Models}\label{subsec:GSM}
Related classes of models are the mixed
\ac{GSM} with a probit link function \citep{Royston02,Liu16,Liu17} and the 
mixed version of 
the linear transformation model \citep[described by][]{Hothorn18}.
For modelling time to events for $i=1,\dots,n$ individuals, we let $T_i^*$ be the 
$i$th individual's survival time. Further, we assume independent censoring times denoted by 
$C_i$ such that we only observe $T_i = \min (T_i^*, C_i)$. We define the event 
indicator $D_i = 1_{\{T_i^* < C_i\}}$.
$D_i$ takes value one 
if the event of the $i$th individual is observed. We let
$S(t\mid \vec x, \vec z, \vec u) = \Prob(T^* > t \mid \vec x, \vec z, \vec u)$
denote the conditional survival function given the fixed effect covariates, 
$\vec x$, the random effect covariates, $\vec z$, and the random effect, 
$\vec u$. 

In a \ac{GSM} the conditional survival function, 
$S(t\mid\vec x,\vec z, \vec u)$, is modelled on some 
link scale defined by a link function $g$. We assume that $g = -\Phi^{-1}$ such that %
$$-\Phi^{-1}(S(t\mid\vec x, \vec z, \vec u)) = 
  -\Phi^{-1}(S_0(t)) + \vec x^\top\vec\beta
  + \vec z^\top\vec u$$%
for some baseline survival function $S_0$. While the above form is 
instructive, we typically model 
$-\Phi^{-1}(S_0(t))$ by a dot product between some spline 
basis over time and a coefficient vector. This can be generalized to %
$$-\Phi^{-1}(S(t\mid\vec x, \vec z, \vec u)) = 
  \vec x^\top(t)\vec\beta
  + \vec z^\top\vec u$$%
to allow for some time-varying design vector $\vec x(t)\in\mathbb{R}^{l}$,
i.e.\ to allow for 
time-varying fixed effects. Though, it is important to 
keep in mind that a monotonicity constraint should be placed on
$\vec x(t)^\top\vec\beta$ since the survival probability needs to 
decrease as a function of time. This often is not an issue when there is 
only a time-varying intercept in which case the constraint can be handled
e.g.\ with an I-splines or a relaxation. 

In order to show the complete data likelihood, we let 
$\mathcal{C} = \{i\in \{1,\dots,n\}:\, d_i = 0\}$
denote the set of indices of the individuals with a censored event time. Similarly, let
$\mathcal{O} = \{1,\dots,n\}\setminus\mathcal{C}$
denote the set of indices of the individuals with an observed event time. 
Further, let $\vec x_j'(t) = (x_{j1}'(t), \dots, x_{jl}'(t))^\top$ denote the 
derivative of the fixed effect covariates with respect to time and
$\mat X^o(\vec T^o) = (\vec x_j(T_j))_{j\in\mathcal{O}}^\top$
be the design matrix for the individuals with an observed event time and
similarly define 
$\mat X^{\prime o}(\vec T^o) = (\vec x_j'(T_j))_{j\in\mathcal{O}}^\top$, 
$\mat Z^o$, $\vec T^o$, 
$\mat X^c(\vec T^c)$, $\mat Z^c$, and $\vec T^c$
where the latter three are for censored individuals. Lastly, 
let $n_o = \lvert\mathcal{O}\rvert$ denote the number of observed 
events and $n_c = \lvert\mathcal{C}\rvert$ denote the number of 
censored events. 
Then the complete data likelihood is %
\begin{equation}
  \phi^{(K)}(\vec u; \vec 0, \mat\Sigma)
  c(\vec t^o, \mat X^o, \vec\beta)
  \phi^{(n_o)}(-\mat X^o(\vec t^o)\vec\beta - \mat Z^o\vec u)
  \Phi^{(n_c)}(-\mat X^c(\vec t^c)\vec\beta - \mat Z^c\vec u) 
  \label{eqn:cmpLLGSM}
\end{equation}%
where 
$c(\vec t^o, \mat X^o, \vec\beta) = -\mat X^{\prime o}(\vec t^o)\vec\beta$. 
Further,
$c(\vec t^o, \mat X^o, \vec\beta) = 1$ and 
$\phi^{(n_o)}(-\mat X^o(\vec t^o)\vec\beta - \mat Z^o\vec u) = 1$ if $n_o = 0$ 
and $\Phi^{(n_c)}(-\mat X^c(\vec t^c)\vec\beta - \mat Z^c\vec u) = 1$ if $n_c = 0$ 
by definition.
We can simplify the log marginal likelihood, 
$l(\vec\beta, \mat\Sigma)$, 
first by defining %
\begin{align*}
\mat H(\vec t^o, \mat Z^o, \mat\Sigma) &= 
  \mat H = 
  \mat Z^{o\top}\mat Z^o + \mat\Sigma^{-1} \\
\vec h(\vec t^o, \mat X^o, \mat Z^o, \mat\Sigma) &=
  \vec h =  
  \mat H^{-1}
  \mat Z^{o\top}\left(- \mat X^o(\vec t^o)\vec\beta
  \right)\\
k(\vec t^o, \mat X^o, \mat Z^o, \mat\Sigma) &= 
  c(\vec t^o, \mat X^o, \vec\beta) 
  (2\pi)^{-n_o/2}\lvert\mat\Sigma\mat H\rvert^{-1/2}  
  \\
&\hspace{20pt}
  \cdot\exp\left( -\frac 12
  (- \mat X^o\vec\beta(\vec t^o))^\top
  (- \mat X^o\vec\beta(\vec t^o)) 
  +\frac 12\vec h^\top\mat H\vec h\right)
\end{align*}%
and we set $\mat H = \mat\Sigma^{-1}$, $\vec h = \vec 0$, and 
$k(\vec t^o, \mat X^o, \mat Z^o, \mat\Sigma) = c(\vec t^o, \mat X^o, \vec\beta)$
if $\mathcal{O} = \emptyset$. 
Then, we can write the log marginal likelihood as %
\begin{equation}\label{eqn:margLLGSM}
l(\vec\beta, \mat\Sigma) 
  = \log k(\vec t^o, \mat X^o, \mat Z^o, \mat\Sigma) +
    \log\int 
    \phi^{(K)}(\vec u; \vec h, \mat H^{-1})
  \Phi^{(n_c)}(-\mat X^c(\vec t^c)\vec\beta - \mat Z^c\vec u)
  \der \vec u
\end{equation}%
which is in the same form as the log marginal likelihood in Equation 
\eqref{eqn:GenericMargLLCDF}. One can observe that a mixed Tobit model is also 
within this class if we set 
$\vec x_j(t) = (-t, x_{j1}, \dots, x_{jl})^\top$ and 
allow for a known offset.
Expressed another way, 
the \ac{GSM} can be seen as a generalization of the Tobit 
model when a probit link is applied.

To apply the \ac{CDF} approach, we can use Equation \eqref{eqn:margLLGSM} 
and Equation \eqref{eqn:margSkewCondRHS} to show that
\begin{align*}
\vec\xi_1 &= \vec h & 
  \vec\xi_2 &= \vec 0 & 
  \mat\Xi_{11} &= \mat H^{-1} & 
  \mat\Xi_{21} &= \mat Z^c\mat H^{-1} \\
  \mat\Xi_{22} &= \mat I + \mat Z^c\mat H^{-1}\mat Z^{c\top} & 
  k_1 &= K & 
  k_2 &= n_c & 
  \vec v_2 &= - \mat X^c(\vec t^c)\vec\beta - \mat Z^c\vec h.
\end{align*}%
Notice that the dimension of the \ac{CDF} in the \ac{CDF} approach is equal 
to the number of censored individuals, $n_c$, making 
this approach very attractive when there are few censored individuals.

We will not show more examples of mixed models for which the 
log marginal likelihood
can be written in the form as for Equation \eqref{eqn:GenericMargLLCDF}. 
We will mention, however, that the joint model of repeated measures 
and time to event data with the discrete hazard survival submodel 
described by \cite{Barrett15}, can be 
derived by using the results from the mixed binomial model in Section 
\ref{subsec:mixProb} in the binary case, $m = 1$, along with similar 
arguments as we use above to get 
from the complete data likelihood in Equation \eqref{eqn:cmpLLGSM} to 
the log marginal likelihood in Equation \eqref{eqn:margLLGSM}.


\section{Simulation Study}\label{sec:SimEx}
We now present a simulation where we study computation times for 
approximating a log marginal likelihood term for the mixed binomial model 
with a single, $m = 1$,  binary outcome. We are particularly interested
in the computation cost as a function of the number of observations, $n$, 
in a cluster and the number of random effect terms, $K$. 
For the calculations, we determine 
a priori the number of nodes to use with (adaptive) \ac{GHQ}, 
and the relative error in the convergence criteria for the \ac{MC} and 
\ac{RQMC} methods
in order to get a comparable precision of the estimates. 
In a practical application this is unknown, but it is important to monitor 
this in order to make a fair comparison. That said, a particular advantage in 
practical applications
of the \ac{MC} method and the \ac{RQMC} methods
that we use is that, unlike \ac{GHQ}, they provide an estimate of the
error.

We performed 100 simulations for each combination of $n$ and $K$. 
We simulated 
the fixed offset, $\mat X\vec\beta$, and the covariance matrix of the random 
effects, $\mat\Sigma$, in each of the 
100 simulations to 
ensure that we test the methods on different data distributions. 
Details are provided in Appendix
\ref{sec:SimExDetails}. 
For all methods we required that the
relative error of the log marginal likelihood estimate 
was expected to be around 0.0006 or less, 
such that we had
three
digits of precision.

\begin{table}[ht]
\centering
\caption{Median computation times in \nth{1000}s of a second for a single cluster. $K$ is the number of random effect terms. For each combination $n$ and $K$, the fastest time is in bold. Some approximation methods failed to get an estimate within the required precision for some simulation samples. This was only the case for some of the non-adaptive version of the methods. Blank cells have not been run because of the long computation time.} 
\label{tbl:compSimMedian}
\begingroup\footnotesize
\begin{tabular}{rrrrrrrr}
  \hline
$n$ & Method/$K$ & 2 & 3 & 4 & 5 & 6 & 7 \\ 
  \hline
 2 & GHQ & 0.035 & 0.034 & 0.034 &  &  &  \\ 
   & AGHQ & 0.033 & 0.030 & 0.030 & 0.033 & 0.032 & 0.030 \\ 
   & CDF & \textbf{0.027} & \textbf{0.025} & \textbf{0.026} & \textbf{0.027} & \textbf{0.025} & \textbf{0.024} \\ 
   & Adaptive Genz \& Monahan (1999) & 0.893 & 5.643 & 1.667 & 1.408 & 3.766 & 2.168 \\ 
   & RQMC & 7.555 & 4.795 & 5.166 &  &  &  \\ 
   & Adaptive RQMC & 6.527 & 15.394 & 10.457 & 7.244 & 11.859 & 7.326 \\ 
   4 & GHQ & 0.061 & 0.285 & 2.228 &  &  &  \\ 
   & AGHQ & \textbf{0.043} & \textbf{0.161} & 0.991 & 0.863 & 0.807 & 0.816 \\ 
   & CDF & 0.698 & 0.412 & \textbf{0.219} & \textbf{0.224} & \textbf{0.203} & \textbf{0.200} \\ 
   & Adaptive Genz \& Monahan (1999) & 2.324 & 6.337 & 24.137 & 27.127 & 29.472 & 22.486 \\ 
   & RQMC & 12.553 & 18.223 & 32.402 &  &  &  \\ 
   & Adaptive RQMC & 8.369 & 9.966 & 15.048 & 14.343 & 9.967 & 11.273 \\ 
   8 & GHQ & 0.130 & 0.900 & 6.463 &  &  &  \\ 
   & AGHQ & \textbf{0.061} & \textbf{0.268} & 1.667 & 5.942 & 52.379 & 217.768 \\ 
   & CDF & 0.922 & 0.639 & \textbf{0.618} & \textbf{0.586} & \textbf{0.588} & \textbf{0.593} \\ 
   & Adaptive Genz \& Monahan (1999) & 1.176 & 5.181 & 24.866 & 51.908 & 58.952 & 104.992 \\ 
   & RQMC & 18.749 & 44.898 & 81.066 &  &  &  \\ 
   & Adaptive RQMC & 7.373 & 8.298 & 11.518 & 14.990 & 19.828 & 27.162 \\ 
  16 & GHQ & 0.371 & 4.020 & 26.136 &  &  &  \\ 
   & AGHQ & \textbf{0.094} & \textbf{0.336} & \textbf{1.736} & 10.452 & 62.796 & 341.623 \\ 
   & CDF & 8.492 & 8.363 & 8.007 & \textbf{8.185} & \textbf{8.042} & \textbf{7.039} \\ 
   & Adaptive Genz \& Monahan (1999) & 0.210 & 1.165 & 9.947 & 30.551 & 23.000 & 59.064 \\ 
   & RQMC & 19.091 & 83.751 & 207.999 &  &  &  \\ 
   & Adaptive RQMC & 3.083 & 5.822 & 7.998 & 11.620 & 11.137 & 16.823 \\ 
  32 & AGHQ & \textbf{0.151} & 0.609 & 3.264 & 19.634 & 118.587 & 629.202 \\ 
   & CDF & 36.997 & 37.496 & 35.886 & 33.027 & 35.338 & 35.556 \\ 
   & Adaptive Genz \& Monahan (1999) & 0.347 & \textbf{0.420} & \textbf{0.874} & \textbf{2.223} & \textbf{1.972} & 8.655 \\ 
   & Adaptive RQMC & 2.358 & 4.027 & 6.707 & 6.057 & 8.520 & \textbf{7.925} \\ 
   \hline
\end{tabular}
\endgroup
\end{table}

\begin{table}[ht]
\centering
\caption{Median computation times in \nth{1000}s of a second for a single cluster for the multinomial data. $K$ is the number of random effect terms and the number of categories. For each combination $n$ and $K$, the fastest time is in bold.} 
\label{tbl:MultcompSimMedian}
\begingroup\footnotesize
\begin{tabular}{rrrrrr}
  \hline
$n$ & Method/$K$ & 3 & 4 & 5 & 6 \\ 
  \hline
 2 & AGHQ & 1.0 & 7.4 & 55.1 & 403 \\ 
   & CDF & \textbf{0.4} & \textbf{0.4} & \textbf{0.6} & \textbf{3} \\ 
   & Adaptive Genz \& Monahan (1999) & 0.6 & 113.5 & 119.8 & 381 \\ 
   & Adaptive RQMC & 21.1 & 72.4 & 70.2 & 124 \\ 
   4 & AGHQ & 2.0 & 14.3 & 106.8 & 812 \\ 
   & CDF & \textbf{0.6} & \textbf{4.8} & 7.6 & \textbf{16} \\ 
   & Adaptive Genz \& Monahan (1999) & 1.2 & 28.6 & \textbf{3.0} & 146 \\ 
   & Adaptive RQMC & 22.2 & 56.3 & 46.0 & 88 \\ 
   8 & AGHQ & 3.6 & 28.0 & 208.7 & 1545 \\ 
   & CDF & 5.8 & 12.1 & 23.0 & 66 \\ 
   & Adaptive Genz \& Monahan (1999) & \textbf{2.2} & \textbf{3.2} & \textbf{5.6} & \textbf{9} \\ 
   & Adaptive RQMC & 6.2 & 23.1 & 17.5 & 44 \\ 
  16 & AGHQ & 4.8 & 27.6 & 169.2 & 1032 \\ 
   & CDF & 13.5 & 35.0 & 80.7 & 186 \\ 
   & Adaptive Genz \& Monahan (1999) & \textbf{4.4} & \textbf{6.6} & \textbf{11.2} & \textbf{17} \\ 
   & Adaptive RQMC & 11.2 & 19.6 & 30.4 & 43 \\ 
  32 & AGHQ & \textbf{5.6} & 24.4 & 113.4 & 452 \\ 
   & CDF & 42.8 & 120.9 & 264.6 & 642 \\ 
   & Adaptive Genz \& Monahan (1999) & 8.8 & \textbf{12.8} & \textbf{22.5} & \textbf{29} \\ 
   & Adaptive RQMC & 21.4 & 37.6 & 59.6 & 72 \\ 
   \hline
\end{tabular}
\endgroup
\end{table}

Table \ref{tbl:compSimMedian} shows the median computation times.
This gives an indication of the time it would take to approximate
a marginal likelihood in seconds 
(obtained by multiplying the cell value by the number of clusters and by
1000), assuming that all clusters have roughly the 
same size and number of cluster specific random effect terms. 
A corresponding table for the mean computation times
(Table \ref{tbl:compSimMean}),
and estimates of the average scaled root mean
square error (Table \ref{tbl:AvgRMSE}) are attached 
in Appendix \ref{sec:SimExDetails}. The data sets are simulated with R's 
default pseudorandom number generator while our implementation
uses Boost's pseudorandom number generator \citep{BoostRandom}.

For both the \ac{GHQ} and the \ac{GWI} approximation 
implemented by \cite{Genz99}, it was preferable to use an 
adaptive procedure in all scenarios but we exclude the non-adaptive version 
of the latter because it was often extremely slow.
The same conclusion applies for the \ac{RQMC} method for the \ac{GWI} 
but there is only a noticeable difference of a larger number of 
observations, $n$. 
The \ac{CDF} approximation was clearly fastest for small 
clusters. However, the \ac{CDF} approach was not preferable
in all cases, as is argued or suggested by 
\cite{Barrett15}. 
For moderate cluster sizes, either the \ac{CDF} or \ac{AGHQ} approaches 
are fastest (depending on the number of random effect terms), and for large 
cluster sizes either \ac{AGHQ} or an adaptive 
version of the \ac{GWI} approximation suggested by \cite{Genz99} or 
\ac{RQMC} are to be  preferred. The former is only 
attractive if $K$ is not too large as otherwise the 
$\bigO{(K^2 + nK)b^K}$ complexity starts to be an issue. 
The use of Equation \eqref{eqn:simplifyGWI} for $K > n$ with the \acp{GWI}
is also clear in Table \ref{tbl:compSimMedian}.

Next, we perform a similar simulation study with the mixed multinomial
model presented in Section \ref{subsec:Multinomial}. We use one random 
effect per category, $K = c$, such that each category has a corresponding
random effect which makes it more or less likely. We provide further details
in Appendix \ref{subsec:MultSimExDetails}.

In this simulation, the \ac{CDF} approach provides an approximation of a 
$n(c-1)$-dimensional integral whereas the \ac{GWI} is $c=K$-dimensional.
However, the integrand approximation we use for the \ac{GWI} requires 
$8n(c - 1)$ evaluations of the standard normal \ac{CDF}
(see Section \ref{subsec:MultIntDetails}). In contrast, the \ac{CDF} 
approximation only requires $nc$ evaluations. However, the \ac{GWI}
based approximations require only $K = c$ evaluations of the inverse standard 
normal \ac{CDF} in contrast to the $n(c- 1)$ evaluations used by the 
\ac{CDF} approximation. The majority of the computation time with all our 
approximations is used on the standard normal \ac{CDF} function or its 
inverse. Thus, this makes the \ac{CDF} method seem attractive based purely 
on a per sample computational cost. However, the integral for the 
\ac{CDF} approach quickly becomes rather high dimensional.

Table \ref{tbl:MultcompSimMedian} shows the median computation times with 
multinomial data. The \ac{CDF} approach is only the fastest option 
when $nc = nK$ is small compared to $K$. 
If this is not true, then \ac{AGHQ} is best when $K = c$ is small and
otherwise
the \ac{MC} method implemented by \cite{Genz99} is the fastest 
option. Lastly, it may be that the \ac{RQMC} method would be the best alternative 
if a higher precision is required because of the \ac{RQMC}
method's convergence rate.

The simulation examples we provide shows that there is not a single 
algorithm, among those that we consider, which is superior for 
all number of the random effect terms, $K$, and number of observations in a 
cluster, $n$, at a fixed precision of the estimate. In fact, in some cases 
there are 
substantial differences in the computation times of the 
different approximations.


\section{Application to Salamander Mating}\label{sec:application}




In this section, we fit a mixed binary 
model with crossed random effects to the salamander mating data
described by \cite{mccullagh89}. 
Crossed random effects models are 
particularly
difficult computationally as the log marginal likelihood typically ends up with 
terms of logs of moderate-to-high dimensional integrals. We will 
compare the \ac{CDF} approximation, the \ac{MC} approximation suggested 
by \cite{Genz99}, and the Laplace approximation. In our 
example, the first-order Laplace approximation is fast but 
yields downward biased estimates of the standard 
deviations of the random effects, as has been observed previously 
\citep{Raudenbush00,Pinheiro06,Joe08}. 

The binary outcome takes value one if the $i$th salamander mating pair is 
successful. Figure 
\ref{fig:sala_fig} in the appendix 
illustrates the design of the study. There are
six clusters, each containing ten males and females which mates with each 
other in such a way that there are 60 attempted mating pairs
in each cluster. Each salamander is either a whiteside or roughbutt breed. 
See \cite{mccullagh89} for further details on the data set. 
We fit a model such that%
\begin{align*}
\vec U_k &= (\vec U_{kf}^\top, \vec U_{km}^\top)^\top 
  \sim N^{(20)}(\vec 0, \diag(\sigma_f^2\mat I, \sigma_m^2\mat I))\\
Y_{ki} \mid \vec U = \vec u
  &\sim \text{Bin}(\Phi(\eta_{ki}), 1) \\
\eta_{ki} &= \beta_0 + \beta_mI_m(k,i) + \beta_fI_f(k,i) \\
&\hspace{20pt} + \beta_{mf}I_f(k,i)I_m(k,i) + u_{kff_{ki}} + u_{kmm_{ki}} 
\end{align*}%
where $\vec U_{km}$ and $\vec U_{kf}$ are, respectively, male and female 
random effects in the $k$th cluster, $I_m(k,i)$ and $I_f(k,i)$ are 
indicators which take value one if, respectively, the male is the whiteside and 
the female is the whiteside, in the $i$th pair in the $k$th cluster, 
and $m_{ki}, f_{ki} \in \{1, \dots, 10\}$ indicates which male and female 
is present in the $i$th pair in the $k$th cluster. See \cite{Pan07} for a 
similar model using \ac{QMC} but with a logit instead of probit link 
function.

The \ac{CDF} version of the log marginal likelihood term for each cluster 
is the log of a 60-dimensional integral, while the 
\ac{GWI} is 20-dimensional. Thus, we do not use \ac{AGHQ} because of the 
$\bigO{(K^2 + nK)b^K}$ complexity. We fitted the model using 
a Laplace approximation using the \texttt{lme4} package 
\citep{Bates15}, using the \ac{CDF} approximation, and using the  approximation
by \cite{Genz98}. The latter two include implemented approximations of
the gradients. Both of our methods were 
first run with a high relative error in the convergence threshold with a 
maximum of 5000 samples for the \ac{GWI} 
approximation and 10000 for the \ac{CDF} 
approximation. We then used a lower 
convergence threshold with a maximum of 25000 samples
and 50000 for the \ac{CDF}.

The total estimation time using the \ac{CDF}
approximation was 
20.5
seconds and the total estimation time of the method by \cite{Genz98} was 
20.2
seconds using 6 threads on a Laptop 
with an Intel\textsuperscript{\textregistered} Core\texttrademark ~i7-8750H 
CPU, on Ubuntu 18.04 with our software compiled with gcc 8.3.0. The Laplace 
approximation was very fast and required only 
0.3 
seconds. Thus, we use the Laplace approximation to get starting values for 
the method which uses the \ac{CDF} approximation and the method by 
\cite{Genz98}. 
We also used MCMC for a Bayesian version of the model 
with a normal distribution prior for each of the slopes with variance 
$10000$ and a inverse gamma prior for each of the random effect variances 
with shape and rate parameter equal to 0.01. The MCMC estimation 
is performed in 
Stata \citep{Stata19} with 120000 samples where 20000 are used as burn-in.
It took 770 seconds to draw the MCMC samples.

\begin{table}[ht]
\centering
\caption{Parameter estimates for the model for the salamander mating data. The MCMC row shows means of the posterior distribution.  The row after the \ac{CDF} and Genz et al.\ method are standard deviations of the estimates when different seeds are used (in parentheses). These are based on 50 different seeds.} 
\label{tbl:SalaEsts}
\begingroup\footnotesize
\begin{tabular}{rrrrrrrrr}
  \hline
 & $\beta_0$ & $\beta_m$ & $\beta_f$ & $\beta_{mf}$ & $\sigma_f$ & $\sigma_m$ & Log-likelihood \\ 
  \hline
CDF & $\hphantom{(}0.612\hphantom{)}$ & $\hphantom{(}-0.425\hphantom{)}$ & $\hphantom{(}-1.707\hphantom{)}$ & $\hphantom{(}2.110\hphantom{)}$ & $\hphantom{(}0.700\hphantom{)}$ & $\hphantom{(}0.670\hphantom{)}$ & $\hphantom{(}-206.877\hphantom{)}$ \\ 
    & $(0.006)$ & $(0.007)$ & $(0.009)$ & $(0.009)$ & $(0.006)$ & $(0.005)$ &  \\ 
  Genz et al. & $\hphantom{(}0.618\hphantom{)}$ & $\hphantom{(}-0.430\hphantom{)}$ & $\hphantom{(}-1.721\hphantom{)}$ & $\hphantom{(}2.128\hphantom{)}$ & $\hphantom{(}0.710\hphantom{)}$ & $\hphantom{(}0.681\hphantom{)}$ & $\hphantom{(}-206.870\hphantom{)}$ \\ 
   & $(0.001)$ & $(0.002)$ & $(0.002)$ & $(0.003)$ & $(0.001)$ & $(0.001)$ &  \\ 
  Laplace & $\hphantom{(}0.601\hphantom{)}$ & $\hphantom{(}-0.419\hphantom{)}$ & $\hphantom{(}-1.731\hphantom{)}$ & $\hphantom{(}2.140\hphantom{)}$ & $\hphantom{(}0.626\hphantom{)}$ & $\hphantom{(}0.576\hphantom{)}$ & $\hphantom{(}-210.107\hphantom{)}$ \\ 
  MCMC means & $\hphantom{(}0.609\hphantom{)}$ & $\hphantom{(}-0.425\hphantom{)}$ & $\hphantom{(}-1.761\hphantom{)}$ & $\hphantom{(}2.177\hphantom{)}$ & $\hphantom{(}0.741\hphantom{)}$ & $\hphantom{(}0.685\hphantom{)}$ &  \\ 
   \hline
\end{tabular}
\endgroup
\end{table}

The estimated parameters are shown in Table \ref{tbl:SalaEsts}. While the 
fixed effects, $\vec\beta$, generally agree, the results suggest a large 
downward bias of the standard deviations with the Laplace approximation.
The maximum likelihood estimates from the \ac{CDF} and method by \cite{Genz98} 
are very similar as expected. However, the \ac{CDF} estimates are less precise
in this case despite the comparable computation time.


\section{Discussion}\label{sec:Conc}
The methods described in this article, for fitting mixed effects models, 
are relevant in a variety of contexts, where practitioners may need guidance 
in choice of computational approach. In biostatistics, two uses of mixed 
models for which computational issues are important are accounting for 
sample relatedness in large genetic/biobank cohorts \citep{Zhou19}
and family-based analysis of genetic and environmental contributions to 
complex survival and categorical outcomes \citep{Pawitan04,Lichtenstein09}. 
Other examples
include crossed random effects such as the mixed model we estimate for the
salamander data set.

We have shown that three different mixed effect models, 
mixed binomial models, mixed multinomial models, and
mixed \acp{GSM}, all with a probit link function, have a similar 
log marginal likelihood where the log marginal likelihood terms, which are
intractable, can
be written as the log of a \ac{GWI} and a multivariate normal \ac{CDF}.

Our simulation example and application show that none of the approximation
methods that we consider is best in all settings. This 
is in contrast with what other authors have argued. 
Particular
methods can be biased or slow in certain settings. Our findings suggests that 
a hybrid procedure where different approximation methods are used for 
each intractable log marginal likelihood term may be optimal. Moreover, our 
simulations provide guidance concerning 
when to approximate the \ac{CDF} and 
when to approximate the \ac{GWI} and, if so, which method to use.

A considerable amount of computation time is spend on evaluating the 
standard normal \ac{CDF} and its inverse with all methods we consider. 
In our experience, using a quicker but less precise method for the 
\ac{CDF} may reduce the computation time for the methods we use with only 
a minor impact on the precision. 

We have only applied the \ac{MC} approximation of the \ac{CDF} suggested by
\cite{Genz92}. Although the examples in \cite{Genz92} and \cite{Hajivassiliou96} 
show that this method is extremely competitive, alternative approximation
methods may be preferable for a low dimensional \ac{CDF}. This is 
particularly true for the bivariate and trivariate 
normal distribution. 
\citet[Section 2.1 and 4.2]{Genz09} provide a number of alternatives. Thus, 
it might be possible to get a faster \ac{CDF} approximation at a fixed 
precision by alternating between different \ac{CDF} approximations.
Furthermore, \cite{Botev17} has recently suggested to add minimax 
tilting to the method suggested by \cite{Genz92} and
shown that the new approach is considerably better
particularly in some higher dimensional scenarios. The new method may expand the 
settings in which the \ac{CDF} approach is more attractive than approximating 
the \ac{GWI}.

Despite that we have compared the \ac{CDF} approach to a number of other 
approaches, our list of alternatives is not exhaustive. 
There are other approaches which 
can be used for approximating the marginal likelihood or for model 
estimation which are interesting. We provide a number of alternatives in 
Appendix \ref{sec:alternatives}. 

\section{Acknolwedgements}
Partial financial support was received from the Swedish e-Science Research 
Center and the Swedish Research Council (2019-00227).
The authors have no relevant financial or non-financial interests to 
disclose.
The software we have written is available at 
https://github.\-com/\-boennecd\-/mixprobit. All the code for the simulations 
and data analysis will be made available.


\appendix

\section{Simulation Study Details}\label{sec:SimExDetails}

\begin{knitrout}
\definecolor{shadecolor}{rgb}{0.969, 0.969, 0.969}\color{fgcolor}\begin{figure}

{\centering \includegraphics[width=\maxwidth]{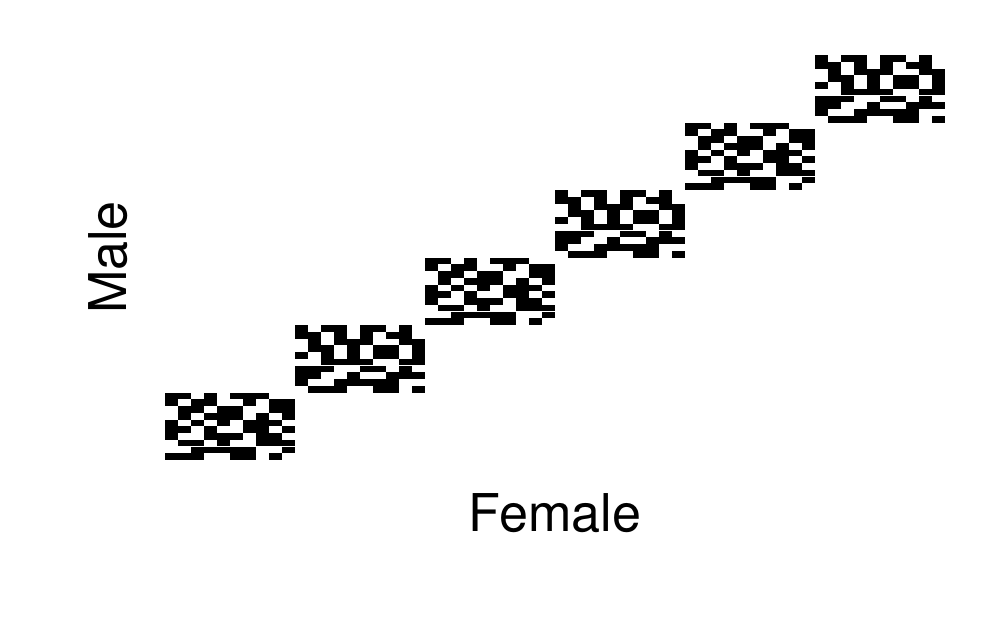} 

}

\caption[Each column represents a female and each row represents a male]{Each column represents a female and each row represents a male. There is a square if a female and male have been mating.}\label{fig:sala_fig}
\end{figure}

\end{knitrout}

For the simulation study in Section \ref{sec:SimEx}, 
the model we simulate from is %
\begin{align*}
\mat\Sigma &\sim W^{(K)}\left(\frac 1{5K} \mat I, 5K\right) 
  & \eta_i = \vec x_i^\top\vec\beta &\sim N(0, 1) \\
\vec U &\sim N^{(K)}(\vec 0, \mat\Sigma)
  & \vec z_i &= \left(K^{-1}, \vec z_i^{\prime\top}\right)^\top \\
\vec z_i' &\sim N^{(K -1)}\left(\vec 0, \frac 1K \mat I\right) 
  & y_i &\sim \text{Bin}(\Phi(\eta_i + \vec z_i^\top\vec u), 1)
\end{align*}%
for $i = 1, \dots, n$ where $W^{(K)}$ denotes a $K$-dimensional 
Wishart distribution.
   
For each of the $n$ and $K$ pairs, we draw 100 samples from 
the above model. The goal is to use different methods 
to approximate the log marginal likelihood 
from taking the log after integrating out the complete data likelihood in 
Equation \eqref{eqn:compL}.
First, we use the importance sampler described in Section 
\ref{sec:SimpleMC}. Let $\hat l^{\text{IS}}$ and $\hat \sigma^{\text{IS}}$
denote the estimate and \ac{MC} standard error estimate, respectively,
from the importance sampler.
Then we start with 
$10 \cdot 10^6$
samples and increase the number of 
samples to a maximum of 
$100 \cdot 10^6$
until %
$$\hat \sigma^{\text{IS}} 4
  < \hat l^{\text{IS}} 0.0002$$%
We fail to reach this precision in very few cases in which case we draw 
new parameters, random effects, and outcomes. 

Next, we ensure the precision of \ac{MC} methods. Let $\hat l^{\text{CDF}}_i$ 
denote the log marginal likelihood estimate for 
$i=1, \dots, 20$ independent approximations 
with the \ac{CDF} approach.
For the \ac{CDF} approach, 
we decrease the relative convergence threshold 
until 
$\sqrt{20^{-1}\sum_{i = 1}^{20}
((\hat l_i^{\text{CDF}} - \hat l^{\text{IS}}) / \hat l^{\text{IS}})^2}$
is less than 0.0002. 
A similar procedure is used for the \ac{GWI} approximation 
suggested by \cite{Genz99} and the \ac{RQMC} method. 
However, we use a maximum of 2500000 samples with all 
three methods.

Typically, it is suggested to increase the number of nodes when using 
quadrature and checking whether this effects the estimates. Thus, we use a 
number of nodes, $b$, such that using
$b - 3, \dots, b$ gives a similar
scaled root mean square error as our \ac{MC} or \ac{RQMC} methods. 
We use a maximum of 
$b = 25$ nodes. All samples where any of the methods 
fail to reach the required precision are excluded. A few samples are 
excluded and those that are is because the non-adaptive methods 
fail to reach the required precision.

The cost of doing the above is not included in the computation times
that we report. The computation times are computed by averaging
over the computation times of 5 runs for each of the 
100 samples given the 
relative error threshold or number of nodes.
\ac{GHQ} and \ac{AGHQ} are implemented in \Cpp . The \ac{GHQ} rule
is implemented with a slightly changed version of the \Cpp 
~implementation in the \texttt{fastGHQuad} package \citep{Blocker18} which 
uses the method suggested by \cite{Golub69}. The 
time used to compute the rules is not included in the times that we report. 
Our \ac{RQMC} method uses the Fortran code 
from the \texttt{randtoolbox} package 
\citep{randtoolbox} which uses the code implemented by \cite{Bratley88} 
and \cite{Joe03}. This has allowed out us to call the Fortran from \Cpp 
~and avoid some large overhead which is present in the R interface of the 
\texttt{randtoolbox} in version 1.30.1. We use 10 
scrambled Sobol sequences in our simulations.

The mode for the adaptive procedures is computed in \Cpp ~with 
the BFGS algorithm 
implemented in the C function \texttt{vmmin} used by R's \texttt{optim} 
function with analytical gradients. The time used to find the mode is 
typically minor compared to the overall cost of the adaptive methods and 
is included in the reported computation times.

\begin{table}[ht]
\centering
\caption{Mean computation times in \nth{1000}s of a second. $K$ is the number of random effect terms. For each combination $n$ and $K$, the fastest time is in bold. Some approximation methods failed to get an estimate within the required precision for some simulation samples. This was only the case for some of the non-adaptive version of the methods. Blank cells have not been run because of the long computation time.} 
\label{tbl:compSimMean}
\begingroup\footnotesize
\begin{tabular}{rrrrrrrr}
  \hline
$n$ & Method/$K$ & 2 & 3 & 4 & 5 & 6 & 7 \\ 
  \hline
 2 & GHQ & 0.039 & 0.038 & 0.045 &  &  &  \\ 
   & AGHQ & 0.034 & 0.031 & 0.031 & 0.034 & 0.033 & 0.031 \\ 
   & CDF & \textbf{0.028} & \textbf{0.026} & \textbf{0.027} & \textbf{0.027} & \textbf{0.026} & \textbf{0.025} \\ 
   & Adaptive Genz \& Monahan (1999) & 39.740 & 32.548 & 23.679 & 18.927 & 20.653 & 19.373 \\ 
   & RQMC & 7.963 & 5.268 & 5.752 &  &  &  \\ 
   & Adaptive RQMC & 56.247 & 68.340 & 49.651 & 25.804 & 38.465 & 32.497 \\ 
   4 & GHQ & 0.065 & 0.349 & 2.669 &  &  &  \\ 
   & AGHQ & \textbf{0.044} & \textbf{0.157} & 0.896 & 1.063 & 0.966 & 1.006 \\ 
   & CDF & 0.836 & 0.592 & \textbf{0.480} & \textbf{0.486} & \textbf{0.410} & \textbf{0.399} \\ 
   & Adaptive Genz \& Monahan (1999) & 35.386 & 44.369 & 109.280 & 95.260 & 89.647 & 114.020 \\ 
   & RQMC & 17.398 & 23.854 & 46.154 &  &  &  \\ 
   & Adaptive RQMC & 82.426 & 85.378 & 59.537 & 60.775 & 36.530 & 42.596 \\ 
   8 & GHQ & 0.156 & 1.187 & 8.660 &  &  &  \\ 
   & AGHQ & \textbf{0.064} & \textbf{0.256} & \textbf{1.475} & 9.065 & 61.779 & 398.004 \\ 
   & CDF & 3.438 & 2.388 & 1.713 & \textbf{1.544} & \textbf{1.118} & \textbf{1.262} \\ 
   & Adaptive Genz \& Monahan (1999) & 45.115 & 40.972 & 157.363 & 174.497 & 230.465 & 308.632 \\ 
   & RQMC & 40.590 & 88.465 & 122.659 &  &  &  \\ 
   & Adaptive RQMC & 117.767 & 92.664 & 44.397 & 48.087 & 60.487 & 99.753 \\ 
  16 & GHQ & 0.414 & 4.687 & 37.815 &  &  &  \\ 
   & AGHQ & \textbf{0.095} & \textbf{0.398} & \textbf{2.167} & 14.551 & 91.090 & 519.403 \\ 
   & CDF & 13.606 & 11.931 & 10.332 & \textbf{9.934} & \textbf{9.456} & \textbf{10.458} \\ 
   & Adaptive Genz \& Monahan (1999) & 10.210 & 24.221 & 137.632 & 253.957 & 237.103 & 251.489 \\ 
   & RQMC & 64.190 & 266.792 & 557.851 &  &  &  \\ 
   & Adaptive RQMC & 28.265 & 41.827 & 21.429 & 23.373 & 25.829 & 26.715 \\ 
  32 & AGHQ & \textbf{0.153} & \textbf{0.615} & \textbf{3.181} & 21.026 & 112.693 & 634.481 \\ 
   & CDF & 67.729 & 66.466 & 75.042 & 64.522 & 59.331 & 46.559 \\ 
   & Adaptive Genz \& Monahan (1999) & 5.165 & 5.177 & 55.343 & 130.605 & 115.364 & 62.633 \\ 
   & Adaptive RQMC & 41.159 & 9.253 & 16.946 & \textbf{13.721} & \textbf{13.643} & \textbf{11.980} \\ 
   \hline
\end{tabular}
\endgroup
\end{table}

\begin{table}[ht]
\centering
\caption{Average scaled root mean square errors (in $10^{-5}$s). $K$ is the number of random effect terms. Some approximation methods failed to get an estimate within the required precision for some simulation samples. This was only the case for some of the non-adaptive version of the methods. Blank cells have not been run because of the long computation time.} 
\label{tbl:AvgRMSE}
\begingroup\footnotesize
\begin{tabular}{rrrrrrrr}
  \hline
$n$ & Method/$K$ & 2 & 3 & 4 & 5 & 6 & 7 \\ 
  \hline
 2 & GHQ & 4.35 & 5.15 & 4.99 &  &  &  \\ 
   & AGHQ & 3.71 & 4.12 & 3.55 & 3.29 & 3.74 & 3.75 \\ 
   & CDF & 0.63 & 0.92 & 0.70 & 0.56 & 0.66 & 0.65 \\ 
   & Adaptive Genz \& Monahan (1999) & 6.90 & 6.77 & 6.64 & 6.98 & 6.60 & 7.54 \\ 
   & RQMC & 7.57 & 7.62 & 7.03 &  &  &  \\ 
   & Adaptive RQMC & 7.32 & 7.73 & 7.39 & 7.50 & 6.88 & 7.69 \\ 
   4 & GHQ & 5.88 & 5.53 & 5.07 &  &  &  \\ 
   & AGHQ & 3.56 & 3.60 & 3.52 & 3.42 & 3.56 & 3.55 \\ 
   & CDF & 7.65 & 7.43 & 7.58 & 8.13 & 7.96 & 7.54 \\ 
   & Adaptive Genz \& Monahan (1999) & 6.44 & 6.85 & 6.89 & 12.11 & 9.94 & 15.06 \\ 
   & RQMC & 7.64 & 7.78 & 7.43 &  &  &  \\ 
   & Adaptive RQMC & 8.21 & 8.11 & 7.02 & 7.55 & 7.43 & 7.62 \\ 
   8 & GHQ & 6.15 & 5.86 & 5.26 &  &  &  \\ 
   & AGHQ & 3.29 & 3.42 & 3.12 & 3.40 & 3.59 & 3.59 \\ 
   & CDF & 7.67 & 7.78 & 7.84 & 8.38 & 7.69 & 7.10 \\ 
   & Adaptive Genz \& Monahan (1999) & 6.24 & 6.49 & 8.30 & 8.14 & 8.56 & 13.89 \\ 
   & RQMC & 7.38 & 7.81 & 8.08 &  &  &  \\ 
   & Adaptive RQMC & 7.04 & 8.22 & 7.22 & 7.39 & 7.21 & 7.48 \\ 
  16 & GHQ & 7.03 & 6.53 & 5.82 &  &  &  \\ 
   & AGHQ & 3.96 & 3.29 & 3.31 & 3.58 & 3.05 & 3.75 \\ 
   & CDF & 6.99 & 8.59 & 6.99 & 6.93 & 7.51 & 7.54 \\ 
   & Adaptive Genz \& Monahan (1999) & 5.51 & 7.63 & 8.37 & 9.00 & 7.85 & 8.55 \\ 
   & RQMC & 7.41 & 7.98 & 8.26 &  &  &  \\ 
   & Adaptive RQMC & 7.26 & 7.54 & 7.80 & 7.21 & 7.62 & 7.25 \\ 
  32 & AGHQ & 4.72 & 4.44 & 4.40 & 3.51 & 3.85 & 3.42 \\ 
   & CDF & 7.41 & 7.18 & 7.44 & 7.78 & 8.30 & 7.43 \\ 
   & Adaptive Genz \& Monahan (1999) & 5.27 & 6.39 & 7.93 & 8.02 & 8.10 & 8.99 \\ 
   & Adaptive RQMC & 7.62 & 7.84 & 7.00 & 7.69 & 8.12 & 7.49 \\ 
   \hline
\end{tabular}
\endgroup
\end{table}

Table \ref{tbl:compSimMean} shows the mean instead of median computation 
times in Table \ref{tbl:compSimMedian}. The discrepancy between the median and
mean shows that we need a lot of nodes or samples for some of the 
log marginal likelihoods that we approximate. 
Table \ref{tbl:AvgRMSE} shows 
the average scaled root mean square error. This is computed as%
$$
\frac 1{100} \sum_{i = 1}^{100}
  \frac
  {\sqrt{5^{-1}
   \sum_{k = 1}^{5}
   \left(\hat l^{\text{CDF}}_{ik} - \hat l^{\text{IS}}_i\right)^2
  }}
  {\hat l^{\text{IS}}_i}
$$%
for the \ac{CDF} approximation where $\hat l^{\text{IS}}_i$ is the 
importance sampler estimate for the $i$th sample and 
$\hat l^{\text{CDF}}_{ik}$ is the $k$th run of the \ac{CDF} approximation
for the $i$th sample. A similar procedure is used for the \ac{MC} methods 
for the \ac{GWI}. The error for the \ac{GHQ} and \ac{AGHQ} is 
computed analogously using 
$b - 3, \dots, b$ nodes.
The latter implies that the \ac{MC} based methods entries' in Table
\ref{tbl:AvgRMSE} are not directly 
comparable with the deterministic \ac{GHQ} and \ac{AGHQ} method.

\subsection{Multinomial Model}\label{subsec:MultSimExDetails}
For the multinomial model we cover in Section \ref{subsec:Multinomial}, 
we include a random effect term for each of the $c$ categories such that 
$K = c$. Each random effect term makes the corresponding category more or 
less likely within the cluster. In particular, the model we simulate from 
is %
\begin{align*}
\mat\Sigma &\sim W^{(K)}\left(\frac 1{5K} \mat I, 5K\right) 
  & \vec \eta_i = \mat B\vec x_i &\sim N^{(c)}(\vec 0, \mat I) \\
\vec U &\sim N^{(K)}(\vec 0, \mat\Sigma)
  & \mat Z_i &= \mat I \\
\mat A_i \mid \vec U = \vec u &\sim N(\mat B\vec x_i + \mat Z_i\vec u, \mat I)
\end{align*}%
with %
$$Y_i = k \Leftrightarrow \forall k \neq k':\, A_{ik} > A_{ik'}, 
  \qquad k,k'\in\{1,\dots,c\}.$$
  
\begin{table}[ht]
\centering
\caption{Mean computation times in \nth{1000}s of a second for the multinomial data. $K$ is the number of random effect terms and the number of categories. For each combination $n$ and $K$, the fastest time is in bold.} 
\label{tbl:MultcompSimMean}
\begingroup\footnotesize
\begin{tabular}{rrrrrr}
  \hline
$n$ & Method/$K$ & 3 & 4 & 5 & 6 \\ 
  \hline
 2 & AGHQ & 1.2 & 9 & 61 & 521 \\ 
   & CDF & \textbf{0.6} & \textbf{1} & \textbf{2} & \textbf{3} \\ 
   & Adaptive Genz \& Monahan (1999) & 19.0 & 345 & 499 & 947 \\ 
   & Adaptive RQMC & 44.2 & 134 & 98 & 181 \\ 
   4 & AGHQ & 2.1 & 16 & 106 & 841 \\ 
   & CDF & \textbf{1.8} & \textbf{5} & \textbf{10} & \textbf{17} \\ 
   & Adaptive Genz \& Monahan (1999) & 30.9 & 368 & 143 & 864 \\ 
   & Adaptive RQMC & 78.8 & 115 & 66 & 122 \\ 
   8 & AGHQ & \textbf{3.2} & 25 & 169 & 1165 \\ 
   & CDF & 6.8 & \textbf{17} & \textbf{32} & 83 \\ 
   & Adaptive Genz \& Monahan (1999) & 11.5 & 84 & 128 & \textbf{70} \\ 
   & Adaptive RQMC & 28.9 & 36 & 34 & 76 \\ 
  16 & AGHQ & \textbf{4.4} & 25 & 167 & 1157 \\ 
   & CDF & 17.8 & 49 & 112 & 263 \\ 
   & Adaptive Genz \& Monahan (1999) & 4.5 & \textbf{7} & \textbf{11} & \textbf{18} \\ 
   & Adaptive RQMC & 12.6 & 22 & 33 & 45 \\ 
  32 & AGHQ & \textbf{5.9} & 25 & 116 & 478 \\ 
   & CDF & 47.4 & 229 & 403 & 1006 \\ 
   & Adaptive Genz \& Monahan (1999) & 8.9 & \textbf{13} & \textbf{23} & \textbf{30} \\ 
   & Adaptive RQMC & 23.2 & 38 & 61 & 72 \\ 
   \hline
\end{tabular}
\endgroup
\end{table}

\begin{table}[ht]
\centering
\caption{Average scaled root mean square errors (in $10^{-5}$s) for the multinomial data. $K$ is the number of random effect terms and the number of categories.} 
\label{tbl:MultAvgRMSE}
\begingroup\footnotesize
\begin{tabular}{rrrrrr}
  \hline
$n$ & Method/$K$ & 3 & 4 & 5 & 6 \\ 
  \hline
 2 & AGHQ & 6.84 & 8.92 & 8.41 & 8.57 \\ 
   & CDF & 19.45 & 19.51 & 21.59 & 20.84 \\ 
   & Adaptive Genz \& Monahan (1999) & 17.06 & 19.85 & 18.37 & 21.65 \\ 
   & Adaptive RQMC & 20.98 & 18.21 & 19.87 & 19.21 \\ 
   4 & AGHQ & 7.02 & 8.00 & 5.81 & 6.60 \\ 
   & CDF & 20.65 & 21.94 & 19.40 & 17.81 \\ 
   & Adaptive Genz \& Monahan (1999) & 13.97 & 19.32 & 18.75 & 18.69 \\ 
   & Adaptive RQMC & 19.08 & 19.28 & 19.24 & 19.95 \\ 
   8 & AGHQ & 7.99 & 7.91 & 7.19 & 7.65 \\ 
   & CDF & 20.73 & 20.90 & 19.91 & 20.79 \\ 
   & Adaptive Genz \& Monahan (1999) & 7.21 & 14.54 & 12.26 & 17.67 \\ 
   & Adaptive RQMC & 18.28 & 20.67 & 19.58 & 20.24 \\ 
  16 & AGHQ & 13.19 & 12.50 & 11.73 & 11.04 \\ 
   & CDF & 20.71 & 18.74 & 19.18 & 18.82 \\ 
   & Adaptive Genz \& Monahan (1999) & 3.16 & 5.25 & 6.28 & 8.85 \\ 
   & Adaptive RQMC & 12.37 & 14.16 & 12.68 & 15.35 \\ 
  32 & AGHQ & 8.53 & 9.39 & 11.98 & 13.17 \\ 
   & CDF & 19.12 & 17.59 & 18.75 & 17.05 \\ 
   & Adaptive Genz \& Monahan (1999) & 1.61 & 2.36 & 1.76 & 2.58 \\ 
   & Adaptive RQMC & 5.23 & 5.88 & 5.88 & 5.61 \\ 
   \hline
\end{tabular}
\endgroup
\end{table}

The other settings are similar to those we use with the mixed binary model 
except that we use a required relative root mean square error of 
0.0005 instead of 0.0002 to reduce
the simulation time. The mean computation times are shown in Table 
\ref{tbl:MultcompSimMean} and average scaled root mean square errors are 
shown in Table \ref{tbl:MultAvgRMSE}. 

\subsubsection{Multinomial Model's Integrand}\label{subsec:MultIntDetails}
The \ac{GWI} for the multinomial has an intractable integrand. That is, 
the conditional probability mass function in Equation 
\eqref{eqn:MultIntegrand} is intractable. We have implemented 
one-dimensional \ac{GHQ} and \ac{AGHQ} to approximate the integrand given 
the random effect, $\vec u$. In particular, for given $\vec\eta$ and 
$\mat K$, we need an approximation of %
$$h(\vec u) = \int
  \phi(a)\Phi^{(c - 1)}(\vec 1a + \vec\eta + \mat K\vec u) \der a,$$%
$\partial/\partial\vec u \log h(\vec u)$, and 
$\partial^2/\partial\vec u\partial\vec u^\top \log h(\vec u)$ where the 
latter two are needed for subsequent adaptive procedures. We perform a 
simulation study where we use the same setup as described in the previous 
section and evaluate the integrand
above at the unknown random effect, $\vec u$. Table 
\ref{tbl:MultIntegrandErr} shows the average absolute relative error and 
Table \ref{tbl:MultIntegrandTime} shows the average evaluation times. The 
adaptive version seems preferable. 
Thus, we have used the \ac{AGHQ} version with 8 nodes given the errors 
shown in Table \ref{tbl:MultIntegrandErr}. 

Moreover, the evaluation times in Table 
\ref{tbl:MultIntegrandTime} shows that the integrand can be approximated 
in a few microseconds. While this is fast, it is an issue as we usually end 
with thousands of evaluations when we approximate the \ac{GWI}.

\begin{table}[ht]
\centering
\caption{Mean absolute relative error times $10^{3}$ from 100 simulations for the integrand in the multinomial model in the \ac{GWI}. The first column shows the number of quadrature nodes.} 
\label{tbl:MultIntegrandErr}
\begingroup\footnotesize
\begin{tabular}{rrrrrrrrrrr}
  \hline
  & $K$ & 2 &  &  & 3 &  &  & 4 &  &  \\ 
   & Method/$n$ & 2 & 4 & 8 & 2 & 4 & 8 & 2 & 4 & 8 \\ 
   \hline
4 & AGHQ & 0.216 & 0.280 & 0.308 & 0.215 & 0.291 & 0.299 & 0.292 & 0.304 & 0.319 \\ 
   & GHQ & 25.521 & 109.780 & 215.267 & 31.981 & 112.965 & 249.999 & 87.333 & 167.709 & 441.312 \\ 
   \\[-.9em]8 & AGHQ & 0.002 & 0.005 & 0.013 & 0.004 & 0.011 & 0.023 & 0.007 & 0.016 & 0.028 \\ 
   & GHQ & 0.517 & 7.053 & 25.589 & 1.312 & 6.463 & 30.082 & 3.613 & 14.542 & 62.799 \\ 
   \\[-.9em]16 & AGHQ & 0.000 & 0.000 & 0.000 & 0.000 & 0.000 & 0.000 & 0.000 & 0.000 & 0.000 \\ 
   & GHQ & 0.001 & 0.079 & 0.261 & 0.005 & 0.121 & 0.658 & 0.030 & 0.225 & 2.447 \\ 
   \\[-.9em]32 & AGHQ & 0.000 & 0.000 & 0.000 & 0.000 & 0.000 & 0.000 & 0.000 & 0.000 & 0.000 \\ 
   & GHQ & 0.000 & 0.000 & 0.001 & 0.000 & 0.000 & 0.002 & 0.000 & 0.000 & 0.009 \\ 
   \hline
\end{tabular}
\endgroup
\end{table}
\begin{table}[ht]
\centering
\caption{Average evaluation times in microseconds from 100 simulations for the integrand in the multinomial model in the \ac{GWI}. The first column shows the number of quadrature nodes.} 
\label{tbl:MultIntegrandTime}
\begingroup\footnotesize
\begin{tabular}{rrrrrrrrrrr}
  \hline
  & $K$ & 2 &  &  & 3 &  &  & 4 &  &  \\ 
   & Method/$n$ & 2 & 4 & 8 & 2 & 4 & 8 & 2 & 4 & 8 \\ 
   \hline
4 & AGHQ & 0.91 & 1.48 & 2.76 & 1.51 & 2.79 & 5.43 & 2.11 & 4.30 & 8.55 \\ 
   & GHQ & 0.62 & 0.99 & 1.81 & 1.05 & 1.87 & 3.57 & 1.46 & 2.84 & 5.35 \\ 
   \\[-.9em]8 & AGHQ & 1.53 & 2.52 & 4.82 & 2.61 & 4.89 & 9.70 & 3.74 & 7.68 & 15.26 \\ 
   & GHQ & 1.24 & 2.00 & 3.65 & 2.06 & 3.78 & 7.21 & 2.91 & 5.74 & 10.93 \\ 
   \\[-.9em]16 & AGHQ & 2.81 & 4.67 & 8.92 & 4.82 & 9.00 & 17.94 & 6.86 & 14.38 & 28.45 \\ 
   & GHQ & 2.48 & 4.05 & 7.46 & 4.16 & 7.57 & 14.58 & 5.88 & 11.80 & 22.22 \\ 
   \\[-.9em]32 & AGHQ & 5.40 & 8.97 & 17.08 & 9.10 & 17.12 & 34.09 & 13.08 & 27.38 & 54.02 \\ 
   & GHQ & 4.94 & 8.20 & 15.19 & 8.36 & 15.34 & 29.66 & 11.81 & 24.05 & 45.58 \\ 
   \hline
\end{tabular}
\endgroup
\end{table}


\section{Alternative Approximations}\label{sec:alternatives}
We describe alternative approximations in this section 
which we have not implemented but 
which have, for particular models, received attention by other authors, 
which we find to be promising but overlooked, or which may be 
useful for practitioners in particular cases.

\subsection{Variational Approximations}\label{subsec:VA}
We will discuss applications of \acp{VA} in this section. We will only give 
a brief introduction to \acp{VA}. See \cite{Bishop06} and \cite{Ormerod10} 
for an introduction to \acp{VA} and examples of 
\acp{VA}.
A typical \ac{VA} uses some family of variational distributions denoted by 
$q$ parameterized by some set 
$\Theta$, which is used in the lower bound of the 
log marginal likelihood. This can be derived by writing the 
log marginal likelihood as%
\begin{align*}
l(\vec\beta, \mat\Sigma) &= \int q(\vec u; \vec\theta)
  \log\left(
  \frac
  {p(\vec y, \vec u)/q(\vec u; \vec\theta)}
  {p(\vec u \mid \vec y)/q(\vec u; \vec\theta)}
  \right)\der\vec u \nonumber\\
  &= 
  \int q(\vec u; \vec\theta)
  \log\left(
  \frac
  {p(\vec y, \vec u)}
  {q(\vec u; \vec\theta)}
  \right)\der\vec u
  + \int q(\vec u; \vec\theta)
  \log\left(
  \frac
  {q(\vec u; \vec\theta)}
  {p(\vec u \mid \vec y)}
  \right)\der\vec u \nonumber
\end{align*}%
Thus%
$$
l(\vec\beta, \mat\Sigma) \geq
  \int q(\vec u; \vec\theta)
  \log\left(
  \frac
  {p(\vec y, \vec u)}
  {q(\vec u; \vec\theta)}
  \right)\der\vec u = \tilde l(\vec\beta, \mat\Sigma, \vec\theta)
$$%
for each $\vec \theta \in \Theta$. The right hand side of the 
in inequality is a lower bound since 
the \ac{KL} divergence %
$$
\int q(\vec u; \vec\theta)
  \log\left(
  \frac
  {q(\vec u; \vec\theta)}
  {p(\vec u \mid \vec y)}
  \right)\der\vec u
$$%
is positive. The idea is to maximize the lower bound,
$\tilde l(\vec\beta, \mat\Sigma, \vec\theta)$, with respect to
$\vec\theta$ to get a tighter lower bound of the log marginal likelihood. 
This is equivalent to minimizing the \ac{KL} divergence between 
$q$ and the conditional density of the random effect, $\vec U$, given the 
observed data, $\vec y$. In particular, if 
$q(\vec u; \vec\theta) = p(\vec u \mid \vec y)$ 
then the lower bound is equal to the log marginal likelihood.
The final \ac{VA} is then given by %
$$
\argmax_{\vec\beta, \mat\Sigma, \vec\theta}
  \tilde l(\vec\beta, \mat\Sigma, \vec\theta)
$$

The important part of a \ac{VA} is to choose a family of variational 
distributions such that it is easy to optimize the lower bound and 
which yields a tight lower bound 
or at least a lower bound that has a similar shape as the log marginal 
likelihood as a function of the model parameters, $\vec\beta$ and 
$\mat\Sigma$. The latter implies that one needs a family of variational 
distributions which has some $\vec\theta\in\Theta$ such that the \ac{KL} 
divergence between the conditional density of the random effects, $\vec U$, 
given the observed data, $\vec y$, is small. 

Two related \acp{VA} are shown for the mixed binomial 
model shown in Section \ref{subsec:mixProb} by \cite{Consonni07} and 
the mixed multinomial model in Section \ref{subsec:Multinomial} by
\cite{Girolami06}. However, the example in \cite{Consonni07} shows that 
such a \ac{VA} may work poorly in certain settings. We now discuss an
alternative \ac{VA} and provide good arguments for why this \ac{VA} may 
work well.

We know that the conditional density of the random effects given the 
observed data is in the of form of the density of a generalized skew-normal 
distribution which density is shown in Equation \eqref{eqn:skewCond}. 
This reduces to the multivariate skew-normal distribution %
$$
q(\vec v_1; \vec\xi_1, \mat\Xi_{11}, \vec\alpha) = 
  2\phi^{(k_1)}(\vec v_1; \vec \xi_1, \mat\Xi_{11})
  \Phi\left(\left(
  \vec\alpha \cdot \diag(\mat \Xi_{11})^{-1/2}
  \right)^\top (\vec v_1 - \vec\xi_1)\right)
$$%
when $k_2 = 1$, $\vec v_2 = \vec 0$, $\vec\xi_2 = \vec 0$, 
$-\mat\Xi_{21}\mat\Xi_{11}^{-1} = (\vec\alpha \cdot \diag(\mat \Xi_{11})^{-1/2})$, 
and $\mat\Xi_{22} - \mat\Xi_{21}\mat\Xi_{11}^{-1}\mat\Xi_{12} = 1$. This is 
interesting as this suggest that the skew-normal \ac{VA} shown by 
\cite{ormerod11} may work well for these models. This is nice since 
\cite{ormerod11} shows that a skew-normal \ac{VA} still yields a quite 
tractable and otherwise easy to approximate lower bound in models like or 
similar to those that we discuss in this paper. Moreover, some of our separate 
work suggest that a skew-normal \ac{VA} works well for mixed \acp{GSM} 
in the sense of providing a very tight lower bound and by being fast to 
optimize. 

If a skew-normal \ac{VA} as suggested in \cite{ormerod11} does not 
yield a tight lower bound, two 
possible generalization of the skew-normal \ac{VA} are %
$$
q(\vec v_1;\vec\xi_1, \mat\Xi_{11},\mat\Lambda) = 
  2^{\tilde k}\phi^{(k_1)}(\vec v_1; \vec \xi_1, \mat\Xi_{11})
  \Phi^{(\tilde k)}\left(
  \mat\Lambda(\vec v_1 - \vec\xi_1)\right)
$$%
or%
$$
q(\vec v_1;\vec\xi_1, \mat\Xi_{11},\vec\lambda, \mat\Lambda, \mat\Xi_{22}) = 
  \phi^{(k_1)}(\vec v_1; \vec \xi_1, \mat\Xi_{11})
  \frac
  {\Phi^{(\tilde k)}\left(
  \vec \lambda
  +\mat\Lambda\mat \Xi_{11}^{-1/2}(\vec v_1 - \vec\xi_1)
  \right)}
  {\Phi^{(\tilde k)}\left(
  \vec \lambda; \vec 0,
  \mat I + \mat\Lambda\mat\Xi_{11}^{-1}\mat\Lambda^\top
  \right)}
$$%
with $1 \leq \tilde k < k_2$, $\mat\Lambda\in\mathbb{R}^{\tilde k \times k_1}$, 
and $\vec\lambda\in\mathbb{R}^{\tilde k}$. 

Using an unconstrained skew-normal \ac{VA} such as suggested by 
\cite{ormerod11} requires estimation of $\bigO{K^2}$-parameters. The 
question then is at what rate one can estimate these parameters. 
A constrained variational distribution can be used 
such that there are only say 
$\bigO{K}$-parameters (e.g.\ by using a diagonal scale matrix). Of course, 
this will result in a less tight lower bound.

To summarize, there are good arguments for using a skew-normal \ac{VA} or 
generalization of it for the class of models that we consider in this paper. 
Despite these arguments, such approximations have received limited 
attention. This is a promising avenue of research for the future, which we 
are currently pursuing, but is beyond the scope of this paper.

\subsection{Expectation Maximization}
A common type of method to estimate mixed models is to use 
an \ac{EM} algorithm 
\citep{Dempster77}. However, the E-step of the \ac{EM} algorithm cannot be
solved analytically in general for the class of models which we consider. 
Thus, some type of quadrature, alternative deterministic procedure, or
\ac{MC} method is needed. Further, the M-step is typically much easier to 
compute if an \ac{ECM} algorithm is used \citep{Meng93}. 

Some authors have suggested \ac{MC} versions of an \ac{ECM} algorithm. 
Two examples are \cite{Hughes99,Vaida07} who develop such algorithms which 
are easily changed to be applicable for the mixed \ac{GSM} model we consider 
in Section \ref{subsec:GSM}. Two disadvantages of such an algorithm are
the sometimes slow convergence of \ac{EM} algorithms,
and that the guarantee that the likelihood is increasing in every iteration 
is lost because of \ac{MC} approximation in the E-step.

\subsection{Hybrid Laplace Approximation}\label{sec:Laplace}
We have already shown the Laplace approximation in Section 
\ref{subsec:Laplace} and provided an application where the estimated 
standard deviations of the random effects are downward biased. 
Interestingly, a hybrid \ac{MC} and Laplace approximation can be considered. 
In particular, the method suggested by \cite{Lai03} can be used for the 
mixed binomial model in Section \ref{subsec:mixProb}. With this method, 
a metric is computed to judge whether or not a Laplace approximation is 
going to give a precise estimate of the log marginal likelihood term for a
given cluster. A \ac{MC} method is used if this is not the case. 

Using such 
a hybrid method in combination with either a \ac{MC} estimator of the 
\ac{GWI}, \ac{AGHQ}, or the \ac{CDF} approximation
depending on the size of the random effects, 
dimension of the \ac{CDF} approximation, and the mean and covariance matrix
can potentially result in a fast and precise method for a large 
group of data sets and class of models. That is, it might be possible to 
develop 
a heuristic for when to use which approximation which is applicable to 
all the mixed models that we discuss in this paper. 
However, it is beyond the scope of
this paper to develop and implement such a hybrid method.

\bibliography{probit-paper}

\begin{thebibliography}{}

\bibitem[\protect\astroncite{Arnold}{2009}]{Arnold09}
Arnold, B.~C. (2009).
\newblock Flexible univariate and multivariate models based on hidden
  truncation.
\newblock {\em Journal of Statistical Planning and Inference}, 139(11):3741 --
  3749.
\newblock Special Issue: The 8th Tartu Conference on Multivariate Statistics \&
  The 6th Conference on Multivariate Distributions with Fixed Marginals.

\bibitem[\protect\astroncite{Azzalini}{2005}]{Azzalini05}
Azzalini, A. (2005).
\newblock The skew-normal distribution and related multivariate families.
\newblock {\em Scandinavian Journal of Statistics}, 32(2):159--188.

\bibitem[\protect\astroncite{Barrett et~al.}{2015}]{Barrett15}
Barrett, J., Diggle, P., Henderson, R., and Taylor-Robinson, D. (2015).
\newblock Joint modelling of repeated measurements and time-to-event outcomes:
  flexible model specification and exact likelihood inference.
\newblock {\em Journal of the Royal Statistical Society: Series B (Statistical
  Methodology)}, 77(1):131--148.

\bibitem[\protect\astroncite{Bates et~al.}{2015}]{Bates15}
Bates, D., M{\"a}chler, M., Bolker, B., and Walker, S. (2015).
\newblock Fitting linear mixed-effects models using {lme4}.
\newblock {\em Journal of Statistical Software}, 67(1):1--48.

\bibitem[\protect\astroncite{Bishop}{2006}]{Bishop06}
Bishop, C.~M. (2006).
\newblock {\em Pattern Recognition and Machine Learning (Information Science
  and Statistics)}.
\newblock Springer-Verlag, Berlin, Heidelberg.

\bibitem[\protect\astroncite{Blocker}{2018}]{Blocker18}
Blocker, A.~W. (2018).
\newblock {\em fastGHQuad: Fast Rcpp Implementation of Gauss-Hermite
  Quadrature}.
\newblock R package version 1.0.

\bibitem[\protect\astroncite{Botev}{2017}]{Botev17}
Botev, Z.~I. (2017).
\newblock The normal law under linear restrictions: simulation and estimation
  via minimax tilting.
\newblock {\em Journal of the Royal Statistical Society: Series B (Statistical
  Methodology)}, 79(1):125--148.

\bibitem[\protect\astroncite{Bratley and Fox}{1988}]{Bratley88}
Bratley, P. and Fox, B.~L. (1988).
\newblock Algorithm 659: Implementing {S}obol’s quasirandom sequence
  generator.
\newblock {\em ACM Trans. Math. Softw.}, 14(1):88–100.

\bibitem[\protect\astroncite{Caflisch}{1998}]{caflisch98}
Caflisch, R.~E. (1998).
\newblock Monte {C}arlo and quasi-{M}onte {C}arlo methods.
\newblock {\em Acta Numerica}, 7:1–49.

\bibitem[\protect\astroncite{Christoffersen et~al.}{2021}]{christoffersen21}
Christoffersen, B., Clements, M., Humphreys, K., and Kjellström, H. (2021).
\newblock Asymptotically exact and fast gaussian copula models for imputation
  of mixed data types.

\bibitem[\protect\astroncite{Christophe and Petr}{2020}]{randtoolbox}
Christophe, D. and Petr, S. (2020).
\newblock {\em randtoolbox: Generating and Testing Random Numbers}.
\newblock R package version 1.30.1.

\bibitem[\protect\astroncite{Consonni and Marin}{2007}]{Consonni07}
Consonni, G. and Marin, J.-M. (2007).
\newblock Mean-field variational approximate {B}ayesian inference for latent
  variable models.
\newblock {\em Computational Statistics \& Data Analysis}, 52(2):790 -- 798.

\bibitem[\protect\astroncite{Cranley and Patterson}{1976}]{Cranley76}
Cranley, R. and Patterson, T. N.~L. (1976).
\newblock Randomization of number theoretic methods for multiple integration.
\newblock {\em SIAM Journal on Numerical Analysis}, 13(6):904--914.

\bibitem[\protect\astroncite{Dempster et~al.}{1977}]{Dempster77}
Dempster, A.~P., Laird, N.~M., and Rubin, D.~B. (1977).
\newblock Maximum likelihood from incomplete data via the {EM} algorithm.
\newblock {\em Journal of the Royal Statistical Society. Series B
  (Methodological)}, 39(1):1--38.

\bibitem[\protect\astroncite{Durbin and Koopman}{1997}]{Durbin97}
Durbin, J. and Koopman, S.~J. (1997).
\newblock Monte {C}arlo maximum likelihood estimation for non-{G}aussian state
  space models.
\newblock {\em Biometrika}, 84(3):669--684.

\bibitem[\protect\astroncite{Genz}{1992}]{Genz92}
Genz, A. (1992).
\newblock Numerical computation of multivariate normal probabilities.
\newblock {\em Journal of Computational and Graphical Statistics},
  1(2):141--149.

\bibitem[\protect\astroncite{Genz and Bretz}{2009}]{Genz09}
Genz, A. and Bretz, F. (2009).
\newblock {\em Computation of Multivariate Normal and t Probabilities}.
\newblock Lecture Notes in Statistics. Springer-Verlag, Heidelberg.

\bibitem[\protect\astroncite{Genz et~al.}{2020}]{mvtnorm}
Genz, A., Bretz, F., Miwa, T., Mi, X., Leisch, F., Scheipl, F., and Hothorn, T.
  (2020).
\newblock {\em {mvtnorm}: Multivariate Normal and t Distributions}.
\newblock R package version 1.0-12.

\bibitem[\protect\astroncite{Genz and Monahan}{1998}]{Genz98}
Genz, A. and Monahan, J. (1998).
\newblock Stochastic integration rules for infinite regions.
\newblock {\em SIAM Journal on Scientific Computing}, 19(2):426--439.

\bibitem[\protect\astroncite{Genz and Monahan}{1999}]{Genz99}
Genz, A. and Monahan, J. (1999).
\newblock A stochastic algorithm for high-dimensional integrals over unbounded
  regions with {G}aussian weight.
\newblock {\em Journal of Computational and Applied Mathematics}, 112(1):71 --
  81.

\bibitem[\protect\astroncite{{Girolami} and {Rogers}}{2006}]{Girolami06}
{Girolami}, M. and {Rogers}, S. (2006).
\newblock Variational {B}ayesian multinomial probit regression with {G}aussian
  process priors.
\newblock {\em Neural Computation}, 18(8):1790--1817.

\bibitem[\protect\astroncite{Golub and Van~Loan}{2013}]{Golub13}
Golub, G.~H. and Van~Loan, C.~F. (2013).
\newblock {\em Matrix Computations}, volume~4.
\newblock JHU Press.

\bibitem[\protect\astroncite{Golub and Welsch}{1969}]{Golub69}
Golub, G.~H. and Welsch, J.~H. (1969).
\newblock Calculation of {G}auss quadrature rules.
\newblock {\em Mathematics of Computation}, 23(106):221--s10.

\bibitem[\protect\astroncite{Guido and Varin}{2012}]{Masarotto12}
Guido and Varin, C. (2012).
\newblock {Gaussian copula marginal regression}.
\newblock {\em Electronic Journal of Statistics}, 6(none):1517 -- 1549.

\bibitem[\protect\astroncite{Hajivassiliou et~al.}{1996}]{Hajivassiliou96}
Hajivassiliou, V., McFadden, D., and Ruud, P. (1996).
\newblock Simulation of multivariate normal rectangle probabilities and their
  derivatives theoretical and computational results.
\newblock {\em Journal of Econometrics}, 72(1):85 -- 134.

\bibitem[\protect\astroncite{Hothorn et~al.}{2018}]{Hothorn18}
Hothorn, T., Möst, L., and Bühlmann, P. (2018).
\newblock Most likely transformations.
\newblock {\em Scandinavian Journal of Statistics}, 45(1):110--134.

\bibitem[\protect\astroncite{Hughes}{1999}]{Hughes99}
Hughes, J.~P. (1999).
\newblock Mixed effects models with censored data with application to {HIV}
  {RNA} levels.
\newblock {\em Biometrics}, 55(2):625--629.

\bibitem[\protect\astroncite{Joe}{2008}]{Joe08}
Joe, H. (2008).
\newblock Accuracy of {L}aplace approximation for discrete response mixed
  models.
\newblock {\em Computational Statistics \& Data Analysis}, 52(12):5066 -- 5074.

\bibitem[\protect\astroncite{Joe and Kuo}{2003}]{Joe03}
Joe, S. and Kuo, F.~Y. (2003).
\newblock Remark on algorithm 659: Implementing {S}obol’s quasirandom
  sequence generator.
\newblock {\em ACM Trans. Math. Softw.}, 29(1):49–57.

\bibitem[\protect\astroncite{Keast}{1973}]{Keast73}
Keast, P. (1973).
\newblock Optimal parameters for multidimensional integration.
\newblock {\em SIAM Journal on Numerical Analysis}, 10(5):831--838.

\bibitem[\protect\astroncite{Lai and Shih}{2003}]{Lai03}
Lai, T.~L. and Shih, M.-C. (2003).
\newblock A hybrid estimator in nonlinear and generalised linear mixed effects
  models.
\newblock {\em Biometrika}, 90(4):859--879.

\bibitem[\protect\astroncite{Lichtenstein et~al.}{2009}]{Lichtenstein09}
Lichtenstein, P., Yip, B.~H., Björk, C., Pawitan, Y., Cannon, T.~D., Sullivan,
  P.~F., and Hultman, C.~M. (2009).
\newblock Common genetic determinants of schizophrenia and bipolar disorder in
  swedish families: a population-based study.
\newblock {\em The Lancet}, 373(9659):234--239.

\bibitem[\protect\astroncite{Lindstrom and Bates}{1990}]{Lindstrom90}
Lindstrom, M.~J. and Bates, D.~M. (1990).
\newblock Nonlinear mixed effects models for repeated measures data.
\newblock {\em Biometrics}, 46(3):673--687.

\bibitem[\protect\astroncite{Liu and Pierce}{1994}]{Liu94}
Liu, Q. and Pierce, D.~A. (1994).
\newblock A note on {G}auss-{H}ermite quadrature.
\newblock {\em Biometrika}, 81(3):624--629.

\bibitem[\protect\astroncite{Liu et~al.}{2016}]{Liu16}
Liu, X.-R., Pawitan, Y., and Clements, M. (2016).
\newblock Parametric and penalized generalized survival models.
\newblock {\em Statistical Methods in Medical Research}, 27(5):1531--1546.
\newblock PMID: 27587596.

\bibitem[\protect\astroncite{Liu et~al.}{2017}]{Liu17}
Liu, X.-R., Pawitan, Y., and Clements, M.~S. (2017).
\newblock Generalized survival models for correlated time-to-event data.
\newblock {\em Statistics in Medicine}, 36(29):4743--4762.

\bibitem[\protect\astroncite{Maurer and Watanabe}{2020}]{BoostRandom}
Maurer, J. and Watanabe, S. (2020).
\newblock {Boost's {C}++ {R}andom library}.

\bibitem[\protect\astroncite{McCullagh and Nelder}{1989}]{mccullagh89}
McCullagh, P. and Nelder, J. (1989).
\newblock {\em Generalized Linear Models, Second Edition}.
\newblock Chapman \& Hall/CRC Monographs on Statistics \& Applied Probability.
  Taylor \& Francis.

\bibitem[\protect\astroncite{McFadden}{1984}]{McFadden84}
McFadden, D.~L. (1984).
\newblock Chapter 24 econometric analysis of qualitative response models.
\newblock In {\em Handbook of Econometrics}, volume~2 of {\em Handbook of
  Econometrics}, pages 1395 -- 1457. Elsevier.

\bibitem[\protect\astroncite{Meng and Rubin}{1993}]{Meng93}
Meng, X.-L. and Rubin, D.~B. (1993).
\newblock Maximum likelihood estimation via the ecm algorithm: A general
  framework.
\newblock {\em Biometrika}, 80(2):267--278.

\bibitem[\protect\astroncite{Morokoff and Caflisch}{1994}]{Morokoff94}
Morokoff, W.~J. and Caflisch, R.~E. (1994).
\newblock Quasi-random sequences and their discrepancies.
\newblock {\em SIAM Journal on Scientific Computing}, 15(6):1251--1279.

\bibitem[\protect\astroncite{Niederreiter}{1972}]{Niederreiter1972}
Niederreiter, H. (1972).
\newblock On a number-theoretical integration method.
\newblock {\em aequationes mathematicae}, 8(3):304--311.

\bibitem[\protect\astroncite{Ochi and Prentice}{1984}]{Ochi84}
Ochi, Y. and Prentice, R.~L. (1984).
\newblock Likelihood inference in a correlated probit regression model.
\newblock {\em Biometrika}, 71(3):531--543.

\bibitem[\protect\astroncite{Ormerod}{2011}]{ormerod11}
Ormerod, J. (2011).
\newblock Skew-normal variational approximations for {B}ayesian inference.
\newblock {\em Unpublished article}.

\bibitem[\protect\astroncite{Ormerod and Wand}{2010}]{Ormerod10}
Ormerod, J.~T. and Wand, M.~P. (2010).
\newblock Explaining variational approximations.
\newblock {\em The American Statistician}, 64(2):140--153.

\bibitem[\protect\astroncite{Owen}{1998}]{Owen98}
Owen, A.~B. (1998).
\newblock Scrambling {S}obol' and {N}iederreiter–{X}ing points.
\newblock {\em Journal of Complexity}, 14(4):466 -- 489.

\bibitem[\protect\astroncite{Pan and Thompson}{2007}]{Pan07}
Pan, J. and Thompson, R. (2007).
\newblock Quasi-{M}onte {C}arlo estimation in generalized linear mixed models.
\newblock {\em Computational Statistics \& Data Analysis}, 51(12):5765 -- 5775.

\bibitem[\protect\astroncite{Pawitan et~al.}{2004}]{Pawitan04}
Pawitan, Y., Reilly, M., Nilsson, E., Cnattingius, S., and Lichtenstein, P.
  (2004).
\newblock Estimation of genetic and environmental factors for binary traits
  using family data.
\newblock {\em Statistics in Medicine}, 23(3):449--465.

\bibitem[\protect\astroncite{Pinheiro and Bates}{1995}]{Pinheiro95}
Pinheiro, J.~C. and Bates, D.~M. (1995).
\newblock Approximations to the log-likelihood function in the nonlinear
  mixed-effects model.
\newblock {\em Journal of Computational and Graphical Statistics}, 4(1):12--35.

\bibitem[\protect\astroncite{Pinheiro and Chao}{2006}]{Pinheiro06}
Pinheiro, J.~C. and Chao, E.~C. (2006).
\newblock Efficient {L}aplacian and adaptive {G}aussian quadrature algorithms
  for multilevel generalized linear mixed models.
\newblock {\em Journal of Computational and Graphical Statistics},
  15(1):58--81.

\bibitem[\protect\astroncite{{R Core Team}}{2019}]{R19}
{R Core Team} (2019).
\newblock {\em R: A Language and Environment for Statistical Computing}.
\newblock R Foundation for Statistical Computing, Vienna, Austria.

\bibitem[\protect\astroncite{Rabe-Hesketh et~al.}{2002}]{Hesketh02}
Rabe-Hesketh, S., Skrondal, A., and Pickles, A. (2002).
\newblock Reliable estimation of generalized linear mixed models using adaptive
  quadrature.
\newblock {\em The Stata Journal}, 2(1):1--21.

\bibitem[\protect\astroncite{Raudenbush et~al.}{2000}]{Raudenbush00}
Raudenbush, S.~W., Yang, M.-L., and Yosef, M. (2000).
\newblock Maximum likelihood for generalized linear models with nested random
  effects via high-order, multivariate {L}aplace approximation.
\newblock {\em Journal of Computational and Graphical Statistics},
  9(1):141--157.

\bibitem[\protect\astroncite{Royston and Parmar}{2002}]{Royston02}
Royston, P. and Parmar, M. K.~B. (2002).
\newblock Flexible parametric proportional-hazards and proportional-odds models
  for censored survival data, with application to prognostic modelling and
  estimation of treatment effects.
\newblock {\em Statistics in Medicine}, 21(15):2175--2197.

\bibitem[\protect\astroncite{{StataCorp.}}{2019}]{Stata19}
{StataCorp.} (2019).
\newblock {\em Stata Statistical Software: Release 16}.
\newblock {StataCorp LLC}, College Station, {TX}.

\bibitem[\protect\astroncite{Vaida et~al.}{2007}]{Vaida07}
Vaida, F., Fitzgerald, A.~P., and DeGruttola, V. (2007).
\newblock Efficient hybrid {EM} for linear and nonlinear mixed effects models
  with censored response.
\newblock {\em Computational Statistics \& Data Analysis}, 51(12):5718 -- 5730.

\bibitem[\protect\astroncite{Wolfinger}{1993}]{Wolfinger93}
Wolfinger, R. (1993).
\newblock Laplace's approximation for nonlinear mixed models.
\newblock {\em Biometrika}, 80(4):791--795.

\bibitem[\protect\astroncite{Zhou et~al.}{2019}]{Zhou19}
Zhou, W., Nielsen, J.~B., Fritsche, L.~G., LeFaive, J., Gagliano~Taliun, S.~A.,
  Bi, W., Gabrielsen, M.~E., Daly, M.~J., Neale, B.~M., Hveem, K., Abecasis,
  G.~R., Willer, C.~J., and Lee, S. (2019).
\newblock Scalable generalized linear mixed model for region-based association
  tests in large biobanks and cohorts.
\newblock {\em bioRxiv}.

\end{thebibliography}

\end{document}